\documentclass[journal]{IEEEtai}
\ifCLASSOPTIONcompsoc
  \usepackage[nocompress]{cite}
\else
  \usepackage{cite}
\fi
\usepackage[colorlinks,urlcolor=blue,linkcolor=blue,citecolor=blue]{hyperref}

\ifCLASSINFOpdf
  \usepackage[pdftex]{graphicx}
  \graphicspath{{fig/}}
\else
\fi
\usepackage{amsmath}
\ifCLASSOPTIONcompsoc
 \usepackage[caption=false,font=footnotesize,labelfont=sf,textfont=sf]{subfig}
\else
 \usepackage[caption=false,font=footnotesize]{subfig}
\fi

\usepackage{booktabs}
\usepackage{xcolor}
\usepackage{amssymb}
\usepackage{bbm}
\usepackage{balance}

\begin{document}
%
\title{Relaxed N-Pairs Loss for Context-Aware Recommendations of Television Content}
%
%
%
%
\author{%
    Miklas~S.~Kristoffersen,
    Sven~E.~Shepstone,
    and~Zheng-Hua~Tan%
\IEEEcompsocitemizethanks{\IEEEcompsocthanksitem This work is supported in part by the Innovation Fund Denmark (IFD) under File No. 5189-00009B. (Corresponding author: Miklas S. Kristoffersen.)
\IEEEcompsocthanksitem M. S. Kristoffersen is with Salling Group A/S, Denmark. This work was done in part while M. S. Kristoffersen was with Bang~\&~Olufsen~A/S and visiting BBC R\&D, London, UK. 
\IEEEcompsocthanksitem S. E. Shepstone is with Bang~\&~Olufsen~A/S, Struer, Denmark. 
\IEEEcompsocthanksitem Z.-H. Tan is with Department of Electronic Systems, Aalborg University, Denmark.}
}

\markboth
{Kristoffersen \MakeLowercase{\textit{et al.}}: Relaxed N-Pairs Loss for Context-Aware
Recommendations of Television Content}
{Kristoffersen \MakeLowercase{\textit{et al.}}: Relaxed N-Pairs Loss for Context-Aware
Recommendations of Television Content}
%



\maketitle

\begin{abstract}
    This paper studies context-aware recommendations in the television domain by proposing a deep learning-based method for learning joint context-content embeddings (JCCE).
    The method builds on recent developments within recommendations using latent representations and deep metric learning, in order to effectively represent contextual settings of viewing situations as well as available content in a shared latent space.
    This embedding space is used for exploring relevant content in various viewing settings by applying an N-pairs loss objective as well as a relaxed variant proposed in this paper.
    Experiments confirm the recommendation ability of JCCE, achieving improvements when compared to state-of-the-art methods.
    Further experiments display useful structures in the learned embeddings that can be used for gaining valuable knowledge of underlying variables in the relationship between contextual settings and content properties.
\end{abstract}

\begin{IEEEImpStatement}
Context-aware recommender systems target content suggestions for users in their current situation. For example, they reduce the decision burden by offering only a few context-adapted suggestions to the user. However, in order to do so, the underlying algorithm must take a variety of heterogeneous features describing the context into account, and it should preferably learn the relationship between these features in a self-supervised manner. The deep learning-based approach, introduced in this paper, jointly models context and content of television viewing in order to improve state-of-the-art results. 

\end{IEEEImpStatement}




%

\section{Introduction}\label{sec:introduction}
%

\IEEEPARstart{R}{ecommender} systems (RS) have become essential for content providers to retain customers by helping them navigate the vast amount of available multimedia.
These systems specialize in delivering personalized suggestions tailored for each individual user through pattern analysis of previous interactions between users and content.
RS is an active field in both industry and academia, and has been so for several years~\cite{Adomavicius2005a,Shi2014}.
A recent development in RS research is the introduction of deep learning-based methods~\cite{Zhang2019}, which have achieved state-of-the-art performance in numerous applications by enabling effective modeling of nontrivial and nonlinear relationships in multimedia consumption.
Deep neural networks have been used to solve complex tasks \cite{Yi2022}, and among others proven useful for learning low-dimensional representations from sparse input data with a large number of attributes \cite{Wang2020}, which is a key challenge in RS with extensive user and content databases.
As an example, context-aware RS (CARS) traditionally use contextual information for pre- or post-filtering recommendations to cater to specific situations, whereas more recent methods model the context directly~\cite{Adomavicius2015,Han2020}.
With the introduction of deep learning-based methods, CARS have seen new exciting opportunities in contextual modeling.
These advancements are mainly driven by the ability to handle heterogeneous information on both context and content, nonlinear feature abstractions in the last layers, as well as reduced efforts for manual feature extraction.
Since many modern RS applications have a large set of features describing context, content, and their interaction, the deep learning-based methods present themselves as a promising tool for studying users' consumption patterns and providing context-relevant recommendations, e.g. based on groups of users in a social setting \cite{Huang2020}.

\begin{figure}[tb]%
    \centering%
    \includegraphics[width=\columnwidth]{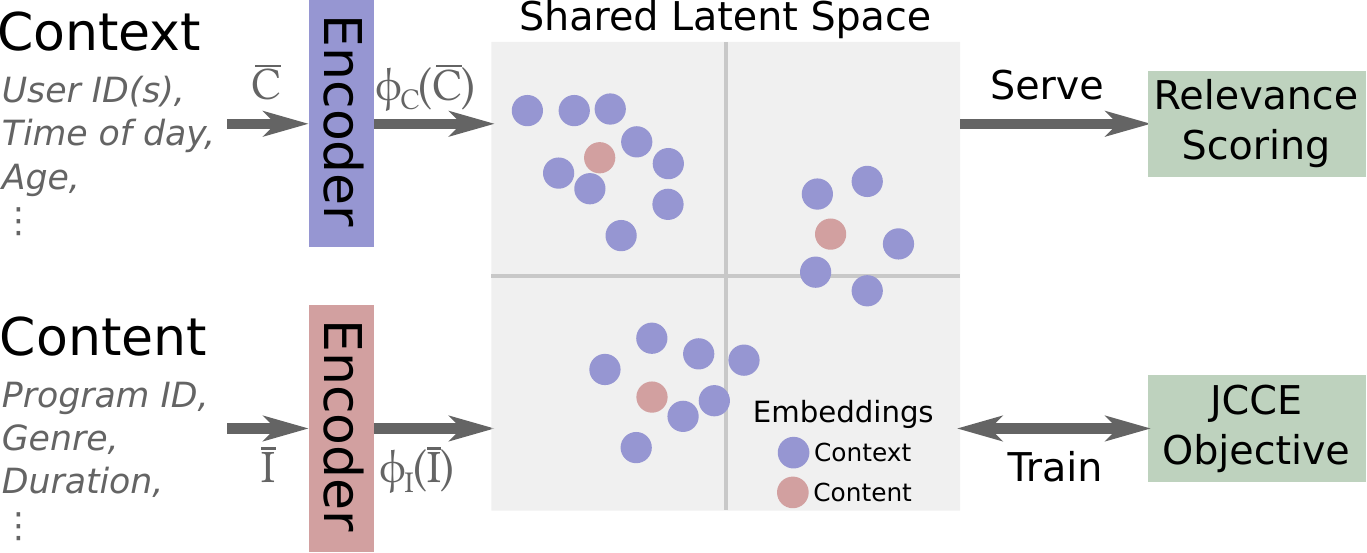}%
\caption{The JCCE model embeds context and content in a shared latent space. During training, observations of context-content interactions are used to jointly train the two encoders with the $N$-pairs loss. For serving content recommendations in a given context, a relevance score is computed for all available items based on similarity in the latent space.}%
\label{fig:fw}%
\end{figure}%

Television is one of the applications in which intelligent adaptation to viewing situations can improve content delivery~\cite{Veras2015}, and where an improved understanding of viewer behavior is important to interested parties, such as content creators, television programmers, and editors.
Previous studies have shown that contextual aspects, e.g. temporal and social settings, have a large impact on viewers' consumption of television, and that recommendations provided in the television domain can be enhanced by utilizing information of viewing situations \cite{Vanattenhoven2015,Kristoffersen2019}.
It is, however, an ongoing challenge for researchers to analyze and recognize the complex interactions among contextual variables and how they link to content preferences.
While some determining aspects of viewing situations can be observed directly and be intuitively included in RS using simple heuristic rules, others are latent compounds that require high-capacity models to uncover.
This motivates the deep learning-based approach investigated in this paper, which, to the best of the authors' knowledge, is the first of its kind within the CARS for television domain.
More specifically, we study how we can provide accurate recommendations based on contextual attributes of viewing situations, and additionally how latent representations can be used for exploring the complex context-content patterns of television consumption.

To address the opportunities and challenges described above, in this paper, we propose the joint context-content embeddings (JCCE) method.
The main idea behind JCCE is the embedding of context and content in a shared latent space as depicted in Fig.~\ref{fig:fw}.
Here context refers to the collection of variables describing a viewing situation, such as temporal aspects and viewer identities, and content represents television content and associated metadata such as genre.
Representing context and content jointly in the same latent space allows us to compute similarities across domains, and thereby ideally enables us to group viewing situations with relevant content.
In addition, the procedure will also group viewing situations that show similar preferences in terms of consumed content, and likewise content that tends to be consumed in similar viewing contexts.
These domain insights can be accessed independently from each other using the respective encoder, and ultimately make the model generalize to unseen context-content pairs.
Since context-content interactions are not explicitly provided by viewers in the form of e.g. ratings, we design JCCE to work with implicit consumption feedback from viewers.
Another advantage of JCCE is that serving recommendations can be done efficiently, since content embeddings only have to be computed once (until we update the model, e.g. with new content).
Thus, to compute recommendations we perform a forward pass of the context encoder and calculate similarities with the precomputed content embeddings, which reduces the number of computations compared to related methods that do not separate the two domains.

The major contributions of the present manuscript are:
\begin{itemize}
    \item This paper extends the methodology descriptions of JCCE and quantitative evaluations of our previous work presented in~\cite{Kristoffersen2019a} with more baseline methods, an additional dataset, and metric.
    \item We propose the relaxed $N$-pairs loss that decreases efforts needed to structure training data as preparation for the learning procedure of JCCE. As described in~\cite{Kristoffersen2019a}, JCCE uses a softmax-based objective with carefully constructed inputs for learning with the $N$-pairs loss. The relaxed loss objective, which we introduce in this paper, enables efficient learning of context-content interactions, and shows improved performance when only few recommendations are allowed.
    \item Furthermore, in addition to the initial results of~\cite{Kristoffersen2019a}, extensive experiments now evaluate, among others, the effect of multiple simultaneous viewers (co-viewing) on recommender performance, and detail both quantitative and qualitative insights from the learned representations in the shared context-content latent space.
\end{itemize}

The outline of the paper is as follows. 
In Section~\ref{sec:rel}, we revisit the existing recommender methods and briefly describe related studies on pattern analysis of contextual aspects in television consumption. 
We then present the proposed JCCE method and the relaxed N -pairs loss in Section~\ref{sec:met}. 
In Section~\ref{sec:exp}, the experimental results on two television datasets and the model performance analyses are given.
Finally, Section~\ref{sec:con} concludes the paper.

\section{Related Work}\label{sec:rel}
We start by giving an overview of findings from using contextual attributes to provide viewers with content recommendations of relevance to their current situation.
Additionally, since our contributions and experiments revolve around deep learning-based methods for recommendations, we describe related research concerned with neural network architectures and loss objectives.

\subsection{Context-Aware Recommendations of Television Content}%
Despite the rapid rise of on-demand services, broadcast television remains one of the main multimedia sources~\cite{EBU2019}, and it is still an economic, cultural, and social important medium, not expected to go away any time soon~\cite{Enli2016,Bruun2020}.
For decades, television has been situated firmly in people's everyday media consumption.
Naturally, researchers have studied numerous methods for helping viewers discover content of interest \cite{Veras2015}.
An early example is presented in \cite{Ardissono2004} which used people meter\footnote{A people meter is a device that automatically logs what is being watched on the television.} data collected from 62 participants over a one year period.
Three components were used: explicit user modeling based on preferences provided by users, stereotypical modeling based on certain demographics of users, and dynamic temporal modeling from implicit feedback inferred from users' viewing behavior.
Recently, \cite{Aharon2015} studied temporal and sequential context, and \cite{Park2017} used temporal aspects for recommending television genres.
In \cite{Hsu2007}, viewers report three mood selections (happy, unhappy, and bored) in addition to temporal information.
A regular feed-forward neural network is used, and the experiments find that information of users' moods improve the performance.
In \cite{Shepstone2014}, users' moods are derived from speech and used to enhance navigation of programs in the electronic program guide.
\cite{Vildjiounaite2009} uses temporal and social context with Support Vector Machines for recommendations. 
They find that social context improves performance depending on how pronounced the family habits are, i.e. the variance in who watches television in the family at certain times of the day.
\cite{Song2012} presents a real-world system that includes information of social context using RFID tags.
The only publicly available people meter dataset in existence (to the best of the authors' knowledge) is presented in \cite{Turrin2014}.
We will later refer to this dataset as the Polimi dataset.
It contains implicit viewing events with timestamps over a four-month period.
In the paper, the authors recommend television content by smoothing temporal context.
A comparable dataset that additionally contains social information is used in \cite{Cremonesi2015}, though the experiments suggest that the added benefit of including social attributes is outweighed by the effect of recommending based on temporal aspects.
\cite{Kristoffersen2019} presents an analysis of the contribution from different contextual aspects to the recommendation performance, e.g. attention level of viewers.
The results indicate that there is a significant gain from including contextual information, and that temporal and social attributes of viewing situations both improve the recommendations.

The research listed above motivates methods for personalization of television content to adapt to specific viewing situations.
As evident, a wide range of contextual attributes have been studied, yet only few methods utilize deep neural networks to model high-level feature abstractions, which have been done successfully in numerous related fields to achieve state-of-the-art performance.
Our work combines CARS for television content with recent advancements in deep learning with the purpose of representing context and content in a latent space suitable for accurate recommendations.

\subsection{Deep and Latent Recommendations}%
Both within RS and CARS neural networks have been investigated as a powerful tool for modeling with large quantities of mostly sparse features.
An early example is presented in~\cite{Salakhutdinov2007a} for learning latent representations of users and content items using Restricted Boltzmann Machines. 
In Prod2Vec~\cite{Grbovic2015}, the RS equivalent to the successful use of embeddings in Word2Vec~\cite{Mikolov2013}, the context of a recommendation is defined from purchase sequences and product side information in the expanded Meta-Prod2Vec~\cite{Vasile2016}.
\cite{Wu2016} similarly relies on advances in natural language processing by providing context-aware recommendations using latent context representations and contextual operating tensors.
In~\cite{Hsieh2017}, a joint user-item space is learned from implicit feedback, and the method integrates item features such as images.
\cite{Covington2016}~presents a real-world CARS implementation of using embeddings for video content recommendation.
In the paper, the concatenated embeddings of various context and content attributes are fed through a single neural network.
DeepCross~\cite{Shan2016} uses a similar architecture, but replaces the standard feed-forward network with a residual network.
Wide \& Deep~\cite{Cheng2016} similarly uses a concatenation of embedded features as input for a feed-forward neural network, but additionally adds a wide component with e.g. cross-product transformed features.
The output is a weighted sum of the wide and deep components.
Factorization Machines~(FM)~\cite{Rendle2010} is a popular method in CARS, which embeds features linearly into a latent space and uses the inner product to model the interactions.
Neural Factorization Machines~(NFM)~\cite{He2017a} combines FM with neural network architectures to effectively model higher-order nonlinear feature interactions. 
In~\cite{Yi2019}, a two-tower architecture is proposed and sampling-bias-corrected softmax is used to learn embeddings from randomly sampled mini-batches. 
\cite{Tanielian2019} presents a relaxed softmax loss using a Boltzmann sampling distribution for negative examples.

A vital component of these systems is the loss objective and the formation of mini-batches with positive and negative interactions.
One of the popular functions is the contrastive loss presented in \cite{Chopra2005}, which minimizes the Euclidean distance between positive embedding pairs while penalizing distances between negative examples smaller than a specified margin.
Another popular approach is the Large Margin Nearest Neighbor (LMNN) formulation in~\cite{Weinberger2006} that samples triplets consisting of a positive embedding pair and a negative pair.
The objective involves separating the positively coupled pair from the negatively coupled pair by a margin, which has inspired the work in e.g. \cite{Hsieh2017}. 
Other variants of the triplet-based loss are introduced in~\cite{Song2016}, where the vector of pairwise distances is lifted to a matrix, and the $N$-pairs loss in \cite{Sohn2016} uses efficient mini-batch construction and generalizes the triplet loss by allowing comparison among multiple negative examples.
In \cite{Opitz2018}, the last embedding layer is divided into an ensemble of embeddings trained using online gradient boosting.

Our work complements these approaches, and takes its starting point in the $N$-pairs loss objective.
We show that the approach may be too constrained, and that relaxing the mini-batch construction constraints positively affects the recommendation performance when we only allow the recommender to present a low number of suggestions.
Within RS, \cite{Yi2019} presents the work closest related to JCCE. 
The methods share a two-tower architecture with a softmax-based learning objective.
\cite{Yi2019} relies on randomly sampled mini-batches with bias correction in the loss, but does not consider that two randomly sampled pairs may not be appropriate as a negative match. 
JCCE handles this situation directly by adopting an $N$-pairs inspired loss objective.

\section{The Proposed JCCE Framework}\label{sec:met}
In this section, we present the JCCE method.
We start by introducing the notation used throughout the paper.
We then motivate JCCE, before we describe how input variables are prepared and then embedded into a shared latent context-content space.
Next, we explain how JCCE serves recommendations, and which objective is used for learning during the training procedure.

The goal of our context-aware television content recommender system is to compose a recommendation list of items sorted by their estimated relevance in a given viewing situation.
Let $\mathcal{V} = [V_1,\ldots,V_L]$ be the log of observed positive interactions between content items and viewing situations, where $L$ is the total number of observations.
Each viewing event specifies the consumption of an item and the associated viewing context, such that $V_i = (I_i,C_i)$, where the item descriptor is a collection of $B$ attributes $I_i=\{I_i^1,\ldots,I_i^B\}$, (as an example, $I_i^1$ can be the program ID of the $i$th viewing event, $I_i^2$ the genre, and so forth).
We let $\mathcal{I}=\{\mathcal{I}_1,\ldots,\mathcal{I}_M\}$ represent the set of distinct items that are available for recommendation, such that $I_i \in \mathcal{I} \text{ for any } i \in \{1,\ldots,L\}$, and denote the cardinality $|\mathcal{I}|=M$.
Viewing situations are described by $D$ context features $C_i=\{C_i^1,\ldots,C_i^D\}$, (as an example, $C_i^1$ can be viewer ID(s) of the $i$th viewing event, $C_i^2$ the weekday, etc.).


A relevance score is estimated for each of the $M$ recommendable items based on the viewing situation $C_i$ 
\begin{equation}
    \hat r_{i,j}=f(C_i,\mathcal{I}_j) \text{ for } j \in  \{1,\ldots,M\},
    \label{eq:rel_score}
\end{equation}
using the trained recommender system, $f$.
The resulting recommendation is a list, $\hat R_i$, of items sorted by relevance to the context $C_i$, such that $\hat R_{i,1}$ and $\hat R_{i,M}$ denote the most and least relevant items, respectively.


The motivation of JCCE is to learn joint embeddings of context and content items as shown in Fig.~\ref{fig:fw}, such that the representations of true pairs obtain high relevance scores, and those pairs that are unlikely to co-occur get low scores -- resulting in early and late placements in the recommendation list, respectively.
As a simplified example, when providing recommendations to a child, children's content is expected to be more preferable than horror movies by having a smaller distance to the context in the latent space.
That is, we estimate relevance of an item from the distance between latent representations of the item and the current context, which means that the learning objective of JCCE revolves around the idea of spatially pulling observed context-content embedding pairs closer together while pushing undesired pairs apart, characteristic to the triplet-based losses~\cite{Weinberger2006}.
However, since the data collection method relies on implicit feedback in the form of observed interactions, negative correlations between content and viewing contexts are generally not present in the data.
As an example, there are no observations to indicate that children should not watch horror movies.
We thus have to carefully consider how we formulate the learning objective and sampling strategy.

\subsection{Preparation and Encoding of Context and Content}\label{sec:met_pre}
Raw examples, $(I_i, C_i)$, can not be used directly in the model, since their features follow different representations, and the encoders require real-valued vectors as input.
Hence, as an initial step, these collections of categorical and numeric features are vectorized. 
For viewing event $V_i$, we denote the vectorized form as $\overline{V}_i=(\overline{I}_i,\overline{C}_i)$ of content item $\overline{I}_i$ and viewing context $\overline{C}_i$, respectively.
Categorical features that contribute with a single value are one-hot encoded, and those contributing with multiple values (e.g. two user IDs) are multi-hot encoded and normalized.
Out-of-vocabulary values (e.g. a new user ID in the test set) are mapped to zero vectors, instead of updating the model which would be the sensible choice in a real-world application.
Numeric features are scaled to $[0,1]$.
The features are then concatenated into a single vector, such that ${\overline{I}_i = [\overline{I_i^1}\,\ldots\,\overline{I_i^B}]}, \overline{I}_i \in \mathbb{R}^{|\overline{I}|}$.
Similarly, ${\overline{C}_i = [\overline{C_i^1}\,\ldots\,\overline{C_i^D}]}$ is used for encoding of the viewing context features. 
%

Let $\phi_I(\overline{I}_i)$ be the embedding of the vectorized content, where $\phi_I: \mathbb{R}^{|\overline{I}|} \rightarrow \mathbb{R}^{E}$ is the content encoding function shown by the red block in Fig.~\ref{fig:fw} and $E$ is the dimensionality of the embedding.
Furthermore, let $\phi_C(\overline{C}_i)$ be the embedding of the vectorized context using $\phi_C: \mathbb{R}^{|\overline{C}|} \rightarrow \mathbb{R}^{E}$, such that the dimensionality of $\phi_I(\overline{I}_i)$ and $\phi_C(\overline{C}_i)$ are identical.
This allows the embeddings to exist in a shared latent space.
As in \cite{Vasile2016}, we set ${E=50}$ based on empirical findings, and do not investigate effects of changing $E$ further. 
We employ nonlinear encoder networks consisting of three fully connected layers each.
The first two layers are each defined to have 250 rectified linear units (ReLUs), and the last layer is a linear transformation with 50 units.
Note that the two encoders can use distinct architectures, but in this work we have opted for similar structures.
Furthermore, previous studies have reported improved performance by using element-wise products to model second-order interactions between embedded features, see e.g.~\cite{He2017a,Beutel2018}.
Such network architectures are also possible to use for the two encoders in JCCE, but are not included in the present contribution.

\subsubsection{Linear JCCE}
Previous studies, e.g. \cite{Mikolov2013,Grbovic2015}, successfully used linear projection.
Therefore, in addition to the encoders described above, we also show the performance of using linear encoders for comparison.
We refer to this method as L-JCCE; A context-aware configuration, which uses the same training objective and procedure as JCCE with linear encoders, $\phi_I(\overline{I}_i) = W_I\overline{I}_i+b_I$ with $W_I\in \mathbb{R}^{E \times |\overline{I}|}$ and $b_I\in \mathbb{R}^{E}$, and similarly $\phi_C(\overline{C}_i) = W_C\overline{C}_i+b_C$ with $W_C\in \mathbb{R}^{E \times |\overline{C}|}$ and $b_C\in \mathbb{R}^{E}$.


\subsection{JCCE Learning Objective and Training Procedure}\label{sec:met_lea}



\begin{figure} [t]
    \centering%
    \subfloat[$N$-Pairs\label{fig:np}]{%
    \includegraphics[width=0.47\columnwidth]{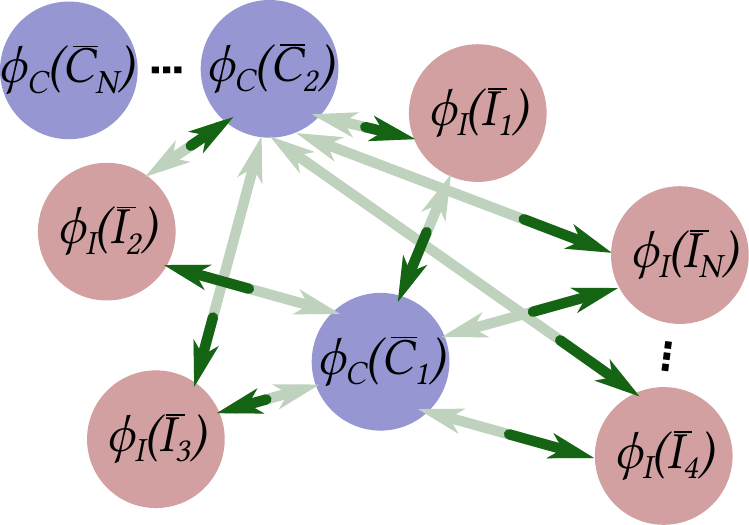}%
    }%
    \hfill
    \subfloat[Relaxed $N$-Pairs\label{fig:rnp}]{%
    \includegraphics[width=0.47\columnwidth]{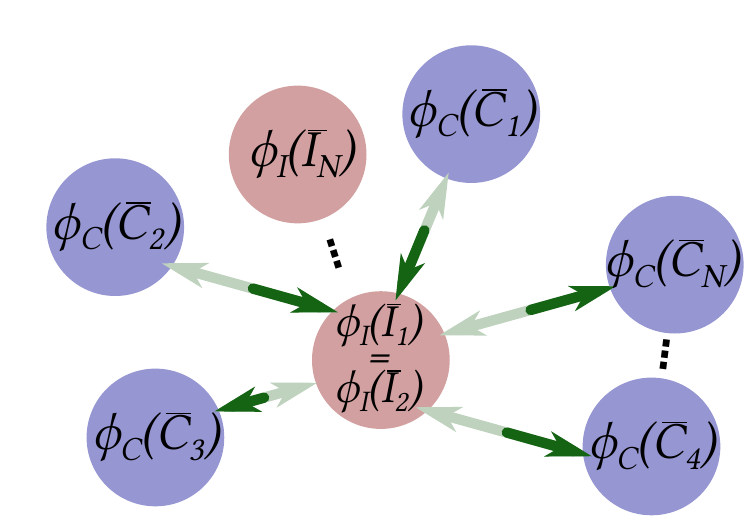}%
    }%
    \caption{Visualization of the push-pull effect of the two losses in the shared context-content latent space. (a) shows how the $N$-pairs loss for each context embedding pulls the associated content embedding while pushing the remaining $N-1$ content embeddings. (b) shows an example of the relaxed $N$-pairs loss with $\phi_I(\overline{I}_1)=\phi_I(\overline{I}_2)$ that pulls the two associated context embeddings while pushing the others.}
\label{fig:npair}
\end{figure}

One approach to reduce distances between observed context-content embeddings, while increasing distances between undesired embedding pairs, is to use a triplet loss~\cite{Schroff2015}.
Within RS, pairwise ranking using e.g. Bayesian Personalized Ranking (BPR)~\cite{Rendle2009} shares the same underlying principles, but is rather motivated from the learning of pairwise relative preferences.
As an example in the context of this paper, consider the embeddings of a viewing context, $C_i$, as an anchor, $a_i$, the associated item, $I_i$, as a positive, $p_i$, and an unobserved item, $I_j\in\mathcal{I}\setminus I_i$, as a negative, $n_i$.
It is then possible to apply e.g. triplet loss and BPR, which would have a high-level objective of minimizing $d(a_i,p_i)-d(a_i,n_i)$, where $d$ is a distance function. 
However, during a single update of the training, these losses only compare an anchor-positive pair to one anchor-negative pair, thereby not including negative examples from the remaining items.
This means that the training step involving $V_i$ solely penalizes one negative item, which often leads to slow convergence.
Hard or semi-hard negative sampling is commonly used to reduce this problem, see e.g.~\cite{Schroff2015}, where the examples that are being mislabeled are used for training, but the selection of these hard examples can be expensive.
Instead, JCCE employs a learning objective which exploits multiple negatives in a single update, as portrayed in Fig.~\ref{fig:npair}.
Anchors, positives, and negatives are often sampled from the same domain, e.g. images of human faces~\cite{Schroff2015}. 
With JCCE we show that similar principles can be used for joint learning of latent viewing context and content representations.


\subsubsection{$N$-Pairs Loss}\label{sec:met_lea_npa}
The $N$-pairs loss was introduced by Sohn in~\cite{Sohn2016} to alleviate the challenges identified above, specifically by introducing an efficient method for constructing mini-batches that allow comparison of anchor-positive pairs to $N-1$ anchor-negative pairs.
In the context of this paper, let $N$ be the number of examples from $\mathcal{V}$ which are included in a single mini-batch update.
From this it follows that $N$ pairs of content items and contexts, $\{(I_i,C_i)\}_{i=1}^N$, are available.
We refer to the vectorized mini-batch of content items as $\overline{I}$ and similarly for viewing contexts we use $\overline{C}$.
By enforcing that the $N$ pairs are examples with $N$ different content types, such that $I_i \neq I_j$ for any $i\neq j$ with $i,j\in \{1,\ldots,N\}$, it is possible to efficiently use examples as both positives and negatives.
That is, as illustrated by the direction of the arrows in Fig.~\ref{fig:np}, for each context example, $C_i$, we can use the associated item, $I_i$, as the only positive and the remaining $N-1$ items as impostors.
Note, however, that the sampling is imperfect, as will be described in Section~\ref{sec:npairs_sampling}, which means that this is a compromised choice, since impostors may not be real impostors.
We then define some probability, $P_i$, that a viewing context, $C_i$, selects content item, $I_i$, as the most relevant target out of all $N$ possibilities.
We use a generalized notation with anchors and positives to allow opposite roles of context and content, and define $P_i$ using softmax with dot product as a measure of similarity between anchor and positive:
\begin{equation}
    P_i=\frac{e^{a_i^\mathsf{T}p_i}}{\sum_{j}^N e^{a_i^\mathsf{T}p_j}},
    \label{eq:np_prob}
\end{equation}
where $a=\{a_i\}_{i=1}^N$ is a collection of $N$ anchor embeddings and $p=\{p_i\}_{i=1}^N$ is a collection of $N$ corresponding positive embeddings.
The $N$-pairs mini-batch construction method makes it simple to compute softmax cross-entropy loss, since each anchor has exactly one positive example:
\begin{align}
    \mathcal{L}_\text{NP}(a,p)=& \frac{1}{N} \sum_{i}^N -\log \left( P_i \right)\nonumber\\
    =& \frac{1}{N} \sum_{i}^N -\log \left( \frac{e^{a_i^\mathsf{T}p_i}}{\sum_{j}^N e^{a_i^\mathsf{T}p_j}}\right)\nonumber\\
    =& \frac{1}{N} \sum_{i}^N \log \left( \sum_{j}^N e^{(a_i^\mathsf{T}p_j - a_i^\mathsf{T}p_i) \mathbbm{1} (i\neq j) } \right) ,
    \label{eq:npairs}
\end{align}
where $\mathbbm{1}$ is an indicator function with a value of one if $i\neq j$ and zero otherwise.
Note that the relevance score used for recommendation in~(\ref{eq:rel_score_cos}) is based only on the direction of the embeddings.
The loss defined in~(\ref{eq:npairs}) is, however, also sensitive to the magnitude of the embeddings, since the vectors used in the dot products are not normalized.
The reason for not applying cosine similarities in place of dot products for the computation of $P_i$ is that the optimization was found empirically to return worse results, confirming the observations in~\cite{Sohn2016}.
Instead, JCCE penalizes large embedding magnitudes through $L_2$ norm regularization:
\begin{equation}
    \mathcal{L}_\text{reg}(a,p)= \lambda \sum_i^N\Vert a_i \Vert ^2 + \Vert p_i \Vert ^2,
\end{equation}
where $\lambda$ is a hyperparameter that controls the regularization strength.
Another solution is to use Euclidean distances in~(\ref{eq:np_prob}), which do not violate the triangle inequality as suggested by Hsieh et al. for a pairwise loss in~\cite{Hsieh2017}.
Instead of regularizing the $L_2$ norm the authors bound the embeddings within the unit circle.\footnote{$L_2$ norm regularization is not applicable in the case with Euclidean distances, since the loss is defined to pull every embedding towards the origin, which does not have a specific meaning in the metric space.}
In practice, we did not obtain notably different results between dot products and Euclidean distances in this work.

Until now we have used context embeddings as anchors and content embeddings as positives in motivating the $N$-pairs loss.
The thought process is equally valid with switched roles of the two, and since the loss is asymmetric by nature, training can be enhanced by including the opposite directed loss \cite{Sohn2016,Jang2018a}.
That is, we define the JCCE training objective as minimizing:
\begin{align}
    \mathcal{L}_\text{JCCE}(\overline{I},\overline{C},\phi_I,\phi_C){}={}& \mathcal{L}_\text{NP}(\phi_I(\overline{I}),\phi_C(\overline{C}))\nonumber\\ & {+} \mathcal{L}_\text{NP}(\phi_C(\overline{C}),\phi_I(\overline{I}))\nonumber\\ & {+} \mathcal{L}_\text{reg}(\phi_I(\overline{I}),\phi_C(\overline{C})).
    \label{eq:jcce_loss}
\end{align}

\subsubsection{$N$-Pairs Sampling}\label{sec:npairs_sampling}
For an ideal $N$-pairs loss the number of pairs in a mini-batch, $N$, equals the total number of distinct content items, $M$, such that each example is compared against all possibilities simultaneously. 
In some tasks, $N=M$ is an acceptable choice that is achieved by randomly sampling a context-content pair for each distinct content, while in others \mbox{$M$-pairs} is too large to fit in memory, forcing us to select a subset ($N<M$), e.g. using negative class mining~\cite{Sohn2016}.

The selection of negative samples is not trivial, since the decision is conditioned on multiple dimensions.
As an example, some content $\mathcal{I}_i$ may not be a negative example in context $C_j$, if $\mathcal{I}_i$ is observed in a very similar context $C_k$ that only differs by being one hour timeshifted from $C_j$.
In the present contribution we use a one-to-one match condition, such that some content $\mathcal{I}_i$ is used as a negative example in context $C_j$ if the pair is not contained in the training set, and keep the exploration of alternative approaches for future work.

\subsubsection{Relaxed $N$-Pairs Loss}\label{sec:met_lea_rnp}
The $N$-pairs loss compares each anchor to one positive and $N-1$ negatives by imposing constraints on the mini-batch construction.
Instead, we propose to trade-off the number of distinct negatives to include multiple positive examples by relaxing the original $N$-pairs loss definition.
Specifically, by loosening the constraints on the mini-batch sampling procedure.
An example is shown in Fig.~\ref{fig:rnp}, where two pairs happen to share the same content, i.e. $I_i=I_j$, thereby motivating us to pull the two associated viewing contexts closer, simultaneously to pushing the remaining negative contexts further away.
The computationally cheapest way to achieve this situation is by relaxing the $N$\nobreakdash-pairs mini-batch sampling to a random selection among all available pairs in $\mathcal{V}$.\footnote{Other more advanced sampling strategies can also be applied, which can potentially improve the performance further, e.g. by sampling from difficult content items or by focusing specifically on learning differences between solitary and social viewing, at the cost of increased computational complexity.}
As a consequence, some content types can be represented multiple times in a mini-batch, while others may not be included at all.
As we will see in the experiments section, performance of the resulting model will vary with the number of recommendations that are allowed to present to viewers.

We denote the indices where the content matches that of $I_i$ as $X_i=\{j|I_i=I_j\}$ including $i=j$, such that $X_i$ has a minimum size, $|X_i|$, of one and a maximum size of $N$.
The probability introduced in (\ref{eq:np_prob}) is then expanded to handle multiple positives.
That is, we define some probability, $P_i'$, that the $i$th anchor selects the $|X_i|$ positive examples over the $N-|X_i|$ negatives:
\begin{equation}
    P_i'
    =\frac{\sum_{j\in X_i} e^{a_i{}^\mathsf{T}p_j}}{\sum_{k}^N e^{a_i{}^\mathsf{T}p_k}}.
\end{equation}
A related idea has been proposed in~\cite{Goldberger2005} for the purpose of Neighborhood Components Analysis (NCA).
The relaxed $N$-pairs loss is different since it selects anchors and positives from two separate domains (it does not compare $a_i$ to $a_j$), while NCA does not distinguish between anchors and positives, but seeks to assign points from the same domain to their nearest labeled neighbor classes.
This distinction is important since it demonstrates that only the $N$-pairs based loss abide the structure of JCCE.
The relaxed $N$-pairs loss is then defined as:
\begin{equation}
    \mathcal{L}_\text{R-NP}(a,p) = \frac{1}{N} \sum_i^N - \frac{1}{|X_i|} \log \left( P_i' \right).
    \label{eq:rnpairs}
\end{equation}
Note that if all the pairs in a mini-batch have distinct content, $|X_i|=1$ for any $i\in \{1,\ldots,N\}$, (\ref{eq:rnpairs}) reduces exactly to the $N$-pairs loss in (\ref{eq:npairs}).
In the opposite case where all pairs in a mini-batch share the same content, $|X_i|=N$ for any $i\in \{1,\ldots,N\}$, the log probability and hence the loss will be zero, though this situation is rare (depending on the sampling strategy, content distribution, and mini-batch size).
In a similar manner as with the regular JCCE, we define the R-JCCE learning objective as minimizing the sum of the two-way relaxed $N$-pairs loss and the weighed embedding magnitudes:
\begin{align}
    \mathcal{L}_\text{R-JCCE}(\overline{I},\overline{C},\phi_I,\phi_C){}={}& \mathcal{L}_\text{R-NP}(\phi_I(\overline{I}),\phi_C(\overline{C}))\nonumber\\ & {+} \mathcal{L}_\text{R-NP}(\phi_C(\overline{C}),\phi_I(\overline{I}))\nonumber\\ & {+} \mathcal{L}_\text{reg}(\phi_I(\overline{I}),\phi_C(\overline{C})).
    \label{eq:r_jcce_loss}
\end{align}


\subsubsection{Training Process}\label{sec:met_lea_train}
We initialize the encoder weights using the uniform Xavier method \cite{Glorot2010}, and initialize the biases as zero.
We use the Adam optimizer~\cite{Kingma2014} with early stopping for learning the encoder weights.  
To reduce the risk of overfitting decisions towards a limited selection of input features (and hence encourage contributions from more context and content variables), we apply dropout~\cite{Srivastava2014}.
The procedure is as follows:
\begin{enumerate}
    \item Sample $N$ pairs from $\mathcal{V}$ using either $N$-pairs sampling or the relaxed counterpart.
    \item Feed the $N$ vectorized context/content samples to the context/content encoder to obtain a total of $2N$ embeddings.
    \item Compute within- and between-pair context-content dot products.
    \item Calculate the loss using (\ref{eq:jcce_loss}) or (\ref{eq:r_jcce_loss}), backpropagate to get the gradient for each weight, and update the encoders using Adam.
    \item Repeat 1-4 until early stopping.
\end{enumerate}
For the experiments, if nothing else is stated, we use a learning rate of 1e-3, dropout rate of 0.1, and regularization strength, $\lambda$, of 5e-4.

\subsection{Serving Recommendations}\label{sec:met_ser}
At serving time, we wish to recommend content based on the contextual settings of specific viewing situations.
That is, if we know the context, $C_i$, we compute a relevance score for some available content, $\mathcal{I}_j$, according to cosine similarity:
\begin{equation}
    \hat r_{i,j} = \frac{\phi_I(\overline{\mathcal{I}}_j)^\mathsf{T}\phi_C(\overline{C}_i)}{\Vert\phi_I(\overline{\mathcal{I}}_j)||\,||\phi_C(\overline{C}_i)||},
    \label{eq:rel_score_cos}
\end{equation}
which is the JCCE solution to~(\ref{eq:rel_score}).
The resulting recommendation list, $\hat R_i$, is the $\hat r_{i,j}$ sorted with decreasing relevance score for the available items.
Note that content embeddings, $\phi_I(\overline{\mathcal{I}}_j)$, are static at serving time, and are only computed once until the model is updated, e.g. with new content.
It is worth highlighting that this makes it cost effective to provide recommendations by performing a forward pass of the context encoder and calculate~(\ref{eq:rel_score_cos}) with the precomputed content embeddings.

\section{Experiments and Validations}\label{sec:exp}
We conduct experiments to evaluate the recommendation and representation ability of JCCE in the television domain.
Using two television datasets (a proprietary and a publicly available) enriched with contextual variables, we show recommendation results of JCCE configurations and compare these to current state-of-the-art methods.
We then present qualitative insights of television consumption through embedding visualizations, and quantitative analyses of entanglement in the learned space.

\subsection{Datasets}\label{sec:exp_dat}
Only few large-scale television consumption datasets exist that include information about viewing contexts.
In this work we use a proprietary dataset that comprises approximately two months of television viewing within the BARB panel.\footnote{The Broadcasters' Audience Research Board (BARB) panel consists of households in the UK. Each household in the panel is equipped with measuring devices connected to each television in the home. These devices are used as a semi-automatic method for participants to report what they are watching when. Participants manually register presence in front of a television using a remote control, which supports multiple simultaneous users. All other variables are collected automatically.}
In the time frame from June to July 2018, approximately 6,000 households encompassing 13,000 unique panel members reported at least once. 
Compared to popular recommender domains, such as music and movies, television content distinguishes itself with e.g. time-constrained and dynamic catalogs~\cite{Turrin2014}.
Therefore, in this work we focus on high-level recommendations of content genres, since these pose as robust descriptors that do not suffer to the same degree as specific programs from the rapidly changing catalog.
In addition to the content genre, we include $D_\text{BARB}=10$ context variables for each viewing event from the BARB data: User IDs, total number of panel members viewing, total number of guests viewing, age distribution of viewers, gender distribution of viewers, day of week, hour of day, television location in the home, UK region, and type of activity (e.g. live or time-shifted). 
We keep viewing events that have a minimum duration of three minutes and remove those that do not, since we can assume that viewers did not engage with the content if they watched it for less.\footnote{A three-minute limit is also used in official reach figures by BARB.}
We furthermore remove viewing events where the consumed content has few total observations ($<1000$), which reduces the number of viewing events from 4 million to 2.7 million.
The reduced BARB dataset contains 94 genres from 13 top-level genres, e.g. \textit{regional} under \textit{news}.
We select the first 90\% of the events for training and the remaining 10\% for evaluation.
The test set covers approximately one week.

We furthermore show results using the publicly available Polimi dataset~\cite{Turrin2014}.
The dataset covers four months in 2013 and has approximately 13,600 active users.
Similarly to the BARB dataset we focus on content genre and include $D_\text{Polimi}=2$ context variables: User ID and hour slot in week.
The hour slot corresponds to the day of week and hour of day variables of the BARB data.
Note, however, that the Polimi dataset only includes a single user ID per viewing event even if multiple users are watching at the same time, while the BARB data include information of multiple simultaneous viewers.
We remove viewing events where the consumed content has few total observations ($<1000$) and events with a duration shorter than three minutes. 
In addition we remove weeks 14 and 19, since they contain errors.
The resulting dataset has a total of 20.5 million viewing events with 99 genres from 8 top-level genres.
Similar to our approach with the BARB dataset, we use the last week for evaluation and the remaining data for training.

\subsection{Baseline Methods}\label{sec:exp_met}

\subsubsection*{Random}\label{sec:exp_met_ran}
Random ranking of content for each viewing event.
A weak baseline that mainly serves as an indicator of coincidental hit chances.
\subsubsection*{Toppop}\label{sec:exp_met_top}
Non-personalized context agnostic ranking of content according to the number of observations in the training set, such that $\hat R_{i,1}$ is the most frequent item in the training set.
It also serves as a measure of dominance among the most popular content compared to that watched less often.
\subsubsection*{Toppop (temp)}\label{sec:exp_met_top2}
The Toppop method is combined with contextual pre-filtering using the temporal settings of viewing events, such that $\hat R_{i,1}$ is the most frequent item within a given hourly slot in the training set.
Temporal context has shown to be a key indicator due to the strong habitual preferences in everyday television consumption~\cite{Turrin2014}.
\subsubsection*{BPR \cite{Rendle2009}}\label{sec:exp_met_bpr}
Bayesian Personalized Ranking (BPR) is a widely used pairwise ranking loss originally designed for matrix factorization.
In this work, we employ the BPR log-sigmoid loss in place of the $N$-pairs loss in JCCE.
This allows BPR to include additional features compared to traditional user-item matrix factorization.
Specifically, we minimize the loss:
\begin{align}
    \mathcal{L}_\text{BPR}(\overline{I},\overline{C},\phi_I,\phi_C) = \sum_i^N & - \log \sigma \left( S(i,i)-S(i,j) \right) \nonumber\\&+ \mathcal{L}_\text{reg}(\phi_I(\overline{I}),\phi_C(\overline{C})),
    \label{eq:bpr}
\end{align}
where $\sigma$ is the sigmoid function, $S$ is a similarity function defined as $S(x,y) = \phi_C(\overline{C}_x)^\mathsf{T}\phi_I(\overline{I}_y)$, and $j$ is a random index of a content embedding in the mini-batch that is not observed together with the $i$th viewing context.

\subsubsection*{Wide \& Deep \cite{Cheng2016}}\label{sec:met_lea_wid}
The Wide \& Deep framework consists of a wide component for memorization and a deep component for generalization.
For both datasets, we use two cross-product transformations for the wide component: 1) user IDs and genre; 2) temporal settings (BARB: Day of week, time of day; Polimi: Hour slot) and genre.
The deep component uses all features and the architecture is chosen to be similar to that of the JCCE encoders.
The model is trained using a logistic loss with observed context-content pairs as positive examples and a similar number of randomly sampled unseen pairs as negative examples.

\subsubsection*{NFM \cite{He2017a}}\label{sec:met_lea_nfm}
NFM combine the linearity of Factorization Machines~\cite{Rendle2010} with the non-linearity of neural networks.
We use an embedding size of 250 factors as input for the bi-interaction layer, and two hidden ReLU layers of 250 and 50 units, respectively.
In a similar manner as Wide \& Deep, we train the NFM model using a logistic loss with observed context-content pairs as positive examples and a similar number of randomly sampled unseen pairs as negative examples.

\subsection{Recommendation Results}\label{sec:exp_rec}
Our first experiment focuses on recommendation performance.
That is, we study to which degree the methods suggest relevant content given unseen viewing situations.
To this end, we define the recommended position $\pi_i$ of observed content $I_i$ in viewing situation $C_i$, such that ${\pi_i = j} | {\hat R _{i,j} = I_i}, {i\in\{1,\ldots,L_\text{test}\}}, {j\in\{1,\ldots,M\}}$, where $L_\text{test}$ is the total number of test samples and $M$ is the number of available items, as previously defined.
Thus, if a method suggests the actual observed content as the most relevant, i.e. $\hat R_{i,1}=I_i$, it follows that $\pi_i=1$.
Worst case $\pi_i=M$.
The recommender performance is then evaluated using hit ratio at K suggestions (HR@K),
\begin{align}
    \text{HR@K} &= \frac{1}{L_\text{test}}\sum_{l=1}^{L_\text{test}} \mathbbm 1 \left( \pi_l \leq \text{K}  \right),
\end{align}
mean reciprocal rank (MRR), 
\begin{align}
    \text{MRR}  &= \frac{1}{L_\text{test}}\sum_{l=1}^{L_\text{test}}\frac{1}{\pi_l},
\end{align}
and area under curve (AUC),
\begin{align}
    \text{AUC}  &= \frac{1}{L_\text{test}\left( M-1 \right) }\sum_{l=1}^{L_\text{test}}M-\pi_l,
\end{align}
where $\mathbbm 1$ is an indicator function with a value of one if $\pi_l$ is less than or equal to K and zero otherwise.

Results for the BARB dataset are listed in Table~\ref{tab:res_barb}.
The Random baseline behaves as expected with a HR@1 of ${1/94}$ and an AUC of 0.5.
The results of Toppop indicate that the most frequently watched genre in the training data is observed in 8\% of the test set viewing, 18.7\% is from the three most popular genres, and on average Toppop places the observed genre at the 17th position in the recommendation list as seen from the AUC ($94-0.832(94-1)\approx17$). 
By including temporal pre-filtering, Toppop (temp) moves the average position of the observed genre three places up the recommendation list to 14th.
The other metrics also improve significantly, thereby confirming that the temporal setting of viewing situations is a simple yet effective feature for improving recommendations.
The remaining methods all outperform the simple Random and Toppop baselines for the BARB dataset.
BPR performs the worst among the six methods in terms of HR@1, HR@3, and MRR, but achieves the third best AUC results.
Note, however, that the AUC results are close for the six methods with a relative improvement of 2.2\% from the worst to the best (Wide \& Deep $\rightarrow$ JCCE).
Correspondingly, HR@1 sees a relative improvement of 61.7\%, HR@3 33.3\%, and MRR 32.0\% (BPR $\rightarrow$ R-JCCE).
Thus, the largest difference in performance between the methods is their ability to position relevant content at the very early placements in the recommendation list. 
BPR, JCCE, and R-JCCE all use identical encoders, but it is clear from the results that the inclusion of multiple negative examples in the training cost function improves performance.
Even with simple linear encoders, L-JCCE outperforms BPR in most metrics.
Wide \& Deep does not achieve scores at the same level as the three best performing methods, possibly due to the training procedure with a point-wise logistic loss compared to the (relaxed) $N$-pairs strategy.
That does, however, not explain the performance gap between Wide \& Deep and NFM, but it confirms the observations in \cite{He2017} which mentions the ability of NFM to model higher-order feature interactions as a contributing factor.
Note that NFM and JCCE achieve almost similar results with the largest difference being HR@3 in favor of NFM.
Both methods outperform R-JCCE in terms of AUC, achieving an average placement of observed genres at 9th and 10th in the recommendation list, respectively.
However, in 33.8\% of the cases R-JCCE recommends the correct genre, compared to 29.3\%, which corresponds to a relative improvement of 15\%.
These results suggest that R-JCCE is strongest with a short recommendation list, while NFM and JCCE are better with more suggestions available.

\begin{table}[t]
\centering%
\caption{Recommendation results for the BARB dataset.}%
\label{tab:res_barb}%
\begin{tabular}{lcccc}
\toprule
         Model &             HR@1$\uparrow$  &             HR@3$\uparrow$  &              MRR$\uparrow$  &              AUC$\uparrow$  \\
\midrule                                                                                     
        Random &            0.011  &            0.032  &            0.055  &            0.500  \\
        Toppop &            0.080  &            0.187  &            0.199  &            0.832  \\
 Toppop (temp) &            0.120  &            0.282  &            0.262  &            0.859  \\
           BPR &            0.209  &            0.414  &            0.363  &            0.906  \\
        L-JCCE &            0.222  &            0.434  &            0.376  &            0.902  \\
  Wide \& Deep &            0.264  &            0.467  &            0.409  &            0.899  \\
           NFM &            0.292  & \underline{0.517} & \underline{0.445} & \underline{0.917} \\
          JCCE & \underline{0.293} &            0.506  &            0.443  &    \textbf{0.919} \\
        R-JCCE &    \textbf{0.338} &    \textbf{0.552} &    \textbf{0.479} &            0.905  \\
\bottomrule
\end{tabular}
\end{table}%
\begin{table}[t]
\centering%
\caption{Recommendation results for the Polimi dataset.}%
\label{tab:res_polimi}%
\begin{tabular}{lcccc}
\toprule
         Model &             HR@1$\uparrow$  &             HR@3$\uparrow$  &              MRR$\uparrow$  &              AUC$\uparrow$  \\
\midrule                                                                                      
        Random &            0.010  &            0.030  &            0.052  &            0.500  \\
        Toppop &            0.145  &            0.351  &            0.302  &            0.885  \\
 Toppop (temp) &            0.201  &            0.444  &            0.375  &            0.924  \\
           BPR &            0.110  &            0.268  &            0.255  &            0.894  \\
        L-JCCE &            0.156  &            0.344  &            0.307  &            0.895  \\
  Wide \& Deep &            0.217  &            0.451  &            0.384  &            0.912  \\
           NFM & \underline{0.259} & \underline{0.524} & \underline{0.438} &    \textbf{0.942} \\
          JCCE &            0.176  &            0.413  &            0.353  & \underline{0.926} \\
        R-JCCE &    \textbf{0.304} &    \textbf{0.529} &    \textbf{0.448} &            0.884  \\
\bottomrule
\end{tabular}
\end{table}%

Results for the Polimi dataset are listed in Table~\ref{tab:res_polimi}.
We have the following key observations.
Firstly, Toppop and Toppop (temp) achieve considerably better scores compared to the BARB dataset.
In fact, the three most popular genres make up more than one third of the viewing, suggesting that the dataset has less diversity in the genres watched by the users.
Combined with a limited number of context parameters available for the context-aware methods, this means that the simple baselines pose as strong contenders for the Polimi dataset.
Secondly, multiple methods struggle with HR@1.
As an example, the HR@1 of JCCE is 40\% lower than that achieved on the BARB dataset, while the AUC actually improves slightly.
Also, despite that more viewing is concentrated around less genres, the HR@1 scores are lower for all six methods compared to the BARB dataset.
This could indicate the advantages of having rich contextual knowledge (as is the case for BARB) when only a very limited number of recommendations are supplied to a user.
Lastly, R-JCCE outperforms NFM for HR@1, HR@3, and MRR, while NFM achieves a better AUC. 

\subsubsection{Effect of Recommendation List Length}\label{sec:exp_rec1}
\begin{figure}[tb]
    \centering
    \includegraphics[width=\columnwidth]{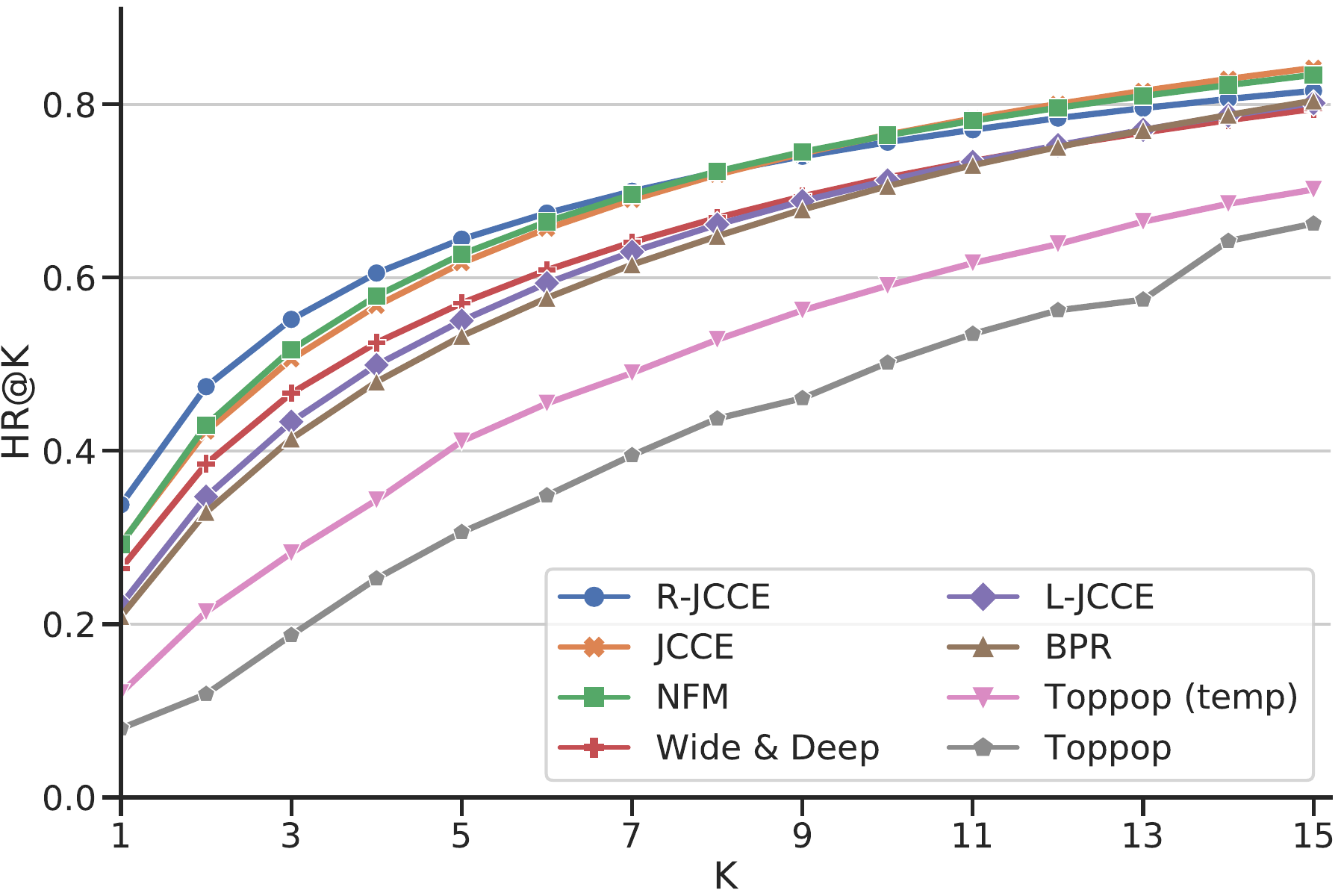}
\caption{HR@K results with different K for the BARB dataset.}
\label{fig:HRatK}
\end{figure}
The recommendation results indicate that some methods perform better with longer recommendation lists, while others are superior with short lists.
In this section, we investigate how the hit ratios of the methods behave at different list lengths. 
Fig.~\ref{fig:HRatK} shows the results (we have not included the Random baseline). 
At low K, R-JCCE performs the best.
As K increases the difference between the methods decreases, which explains the relatively similar AUC observed in Table~\ref{tab:res_barb}.
Note that at K=8, the R-JCCE curve crosses the JCCE curve.
The trade-off between HR @ small K and AUC arises from relaxing the $N$-pairs, which allows JCCE to weigh frequently watched genres more heavily during the training process.
This will put more emphasis on distinguishing popular genres from each other, thereby increasing e.g. the HR@1.
This happens at the expense of less frequent genres' impact while learning.

\subsubsection{Effect of Co-Viewing}\label{sec:exp_rec2}
\begin{figure}[b]
    \centering
    \includegraphics[width=\columnwidth]{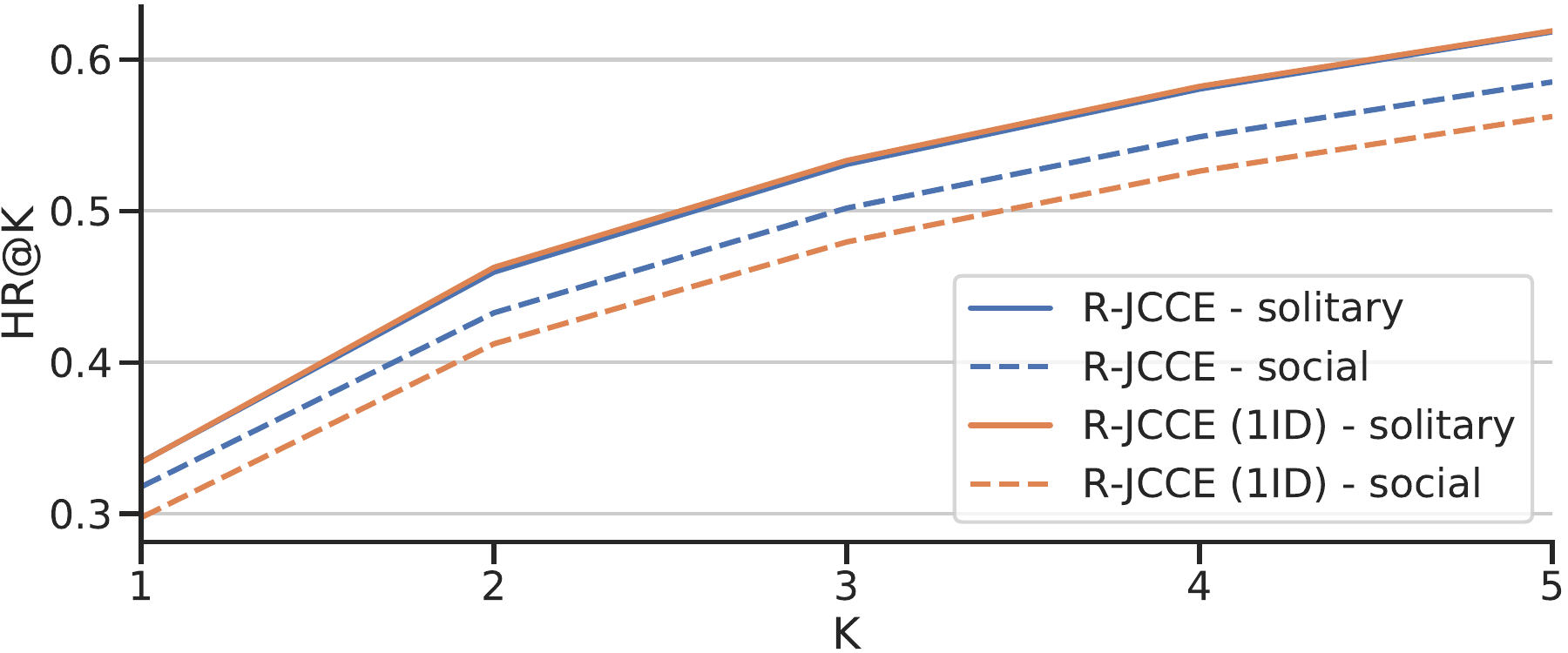}
\caption{HR@K performance at different K, where the color indicates whether the model was trained to be socially aware. The solid lines show the performance in viewing situations with only one viewer, and the dashed lines are for viewing situations with co-viewing.}
\label{fig:HRatK2}
\end{figure}

Previous studies have highlighted social setting as a key feature together with temporal settings in the television RS domain \cite{Vanattenhoven2015,Kristoffersen2019}.
The BARB dataset includes information of multiple simultaneous viewers engaging in a social situation in front of a television, also referred to as users who are co-viewing, and it is therefore of interest to explore the impact of using such information in the JCCE model.
To this end, we train an identical model, but for each viewing situation that has more than one viewer we randomly select one of the viewers and neglect the rest.
We do this to provide a baseline that we can compare against.
We refer to this kind of model as (1ID).
Fig.~\ref{fig:HRatK2} shows how R-JCCE compares to R-JCCE (1ID) in terms of HR@K.
For the comparison we divided the test set into a solitary (70\% of cases) and a social viewing (30\%) set. 
Note that both methods perform better when only one viewer is present, and that their performance is similar in that case.
This is expected, though the (1ID) model could potentially have seen a decrease in performance due to user profile dilution during training from assigning social situations to one specific viewer that may not share the common interest of the group.
When inspecting situations with co-viewing, the original model improves HR@1 by 7\% relative to the (1ID) model, thereby indicating the gain in performance from having information of multiple simultaneous viewers. 

\subsection{Representation Quality}\label{sec:exp_rep}
To further study the ability of JCCE to learn joint context-content representations that achieve high recommendation scores as shown in the previous experiments while also enabling valuable insights to the television domain, we now evaluate the representation quality.
For this purpose, we will use the BARB dataset since it includes a wider range of variables describing the viewing context.

\subsubsection{Content Embeddings}\label{sec:exp_rep1}
\begin{figure}[tb]
    \centering
    \includegraphics[width=\columnwidth]{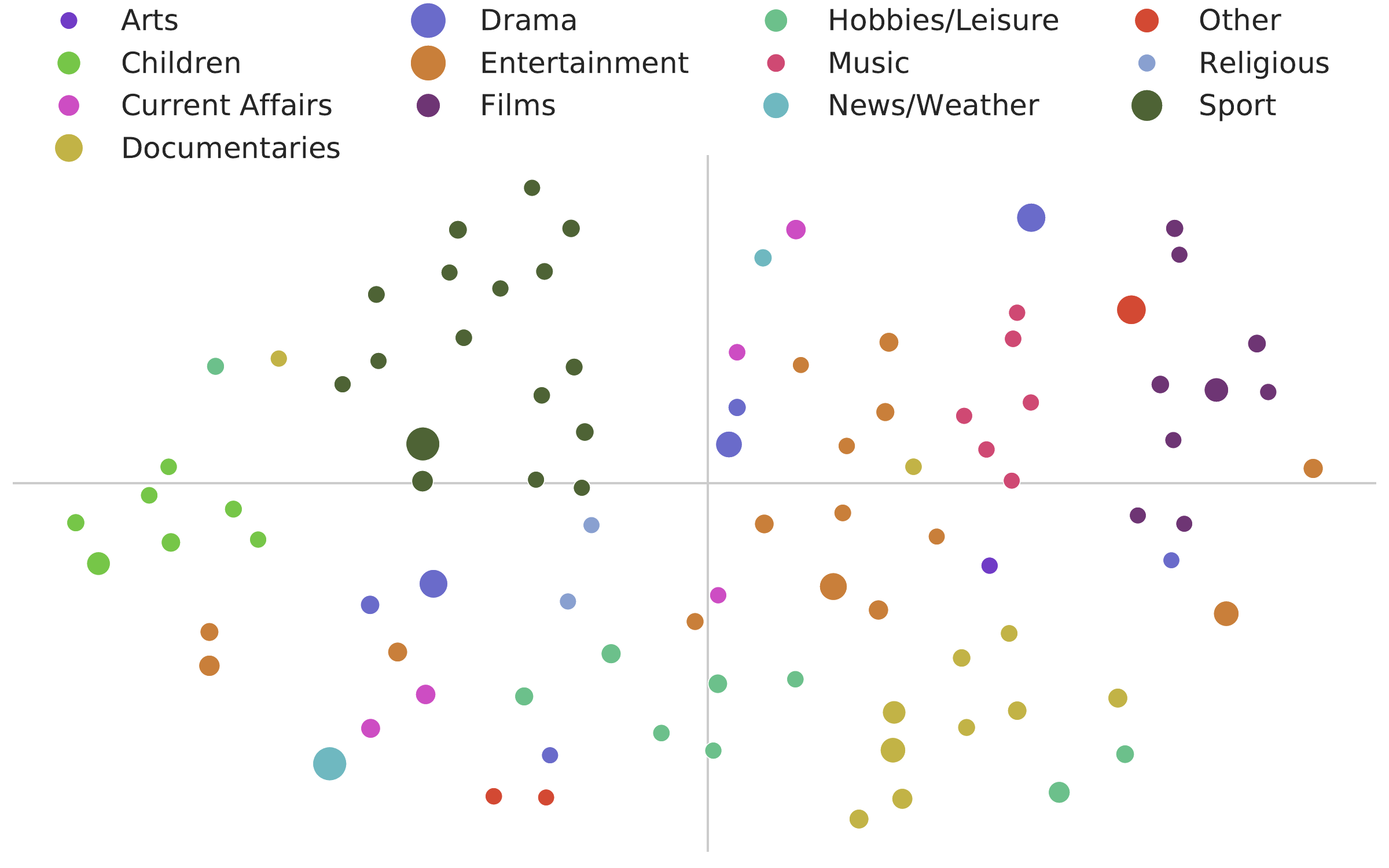}
\caption{BARB genre embeddings of the R-JCCE model reduced to two dimensions using t-SNE. Each point is a sub genre of the top level genre (e.g. \textit{International Football} is a sub genre of \textit{Sport}) scaled by the number of occurrences in the dataset.}
\label{fig:content_emb}
\end{figure}

We start by exploring how the content encoder maps genres to the shared latent space.
To do so, we compute the 50-dimensional embeddings of all 94 genres in the BARB dataset using the content encoder, $\phi_I$.
For the purpose of qualitatively inspecting which genres group together and thus tend to be observed in similar viewing contexts, we use t-SNE \cite{Maaten2008} to reduce the embeddings to two dimensions based on cosine similarity.
Fig.~\ref{fig:content_emb} shows the resulting t-SNE embeddings.
Note how most genres group together under their top level genre, and that this happens without any expert knowledge built into the model.
That is, at no point has the model been informed that e.g. \textit{History} and \textit{Science/Medical} both belong to the top level genre \textit{Documentaries}, but it turns out that R-JCCE learns to map the two genres close to each other in the embedding space.\footnote{This is not directly visible from the figure due to the complexity of visualizing all 94 genres with their labels.}
There are also examples of genre neighbors that do not share the same top level genre, e.g. \textit{Political (Current Affairs)} and \textit{National/International (News)}, and \textit{Classical (Music)} and \textit{Gardening (Hobbies/Leisure)}.
To understand why genres group together the way they do, next we will study the underlying aspects of viewing situations associated with each genre.

\subsubsection{Context Embeddings}\label{sec:exp_rep2}
\begin{figure*}
    \centering%
    \subfloat[Observed genre\label{fig:a}]{
        \includegraphics[width=0.33\textwidth]{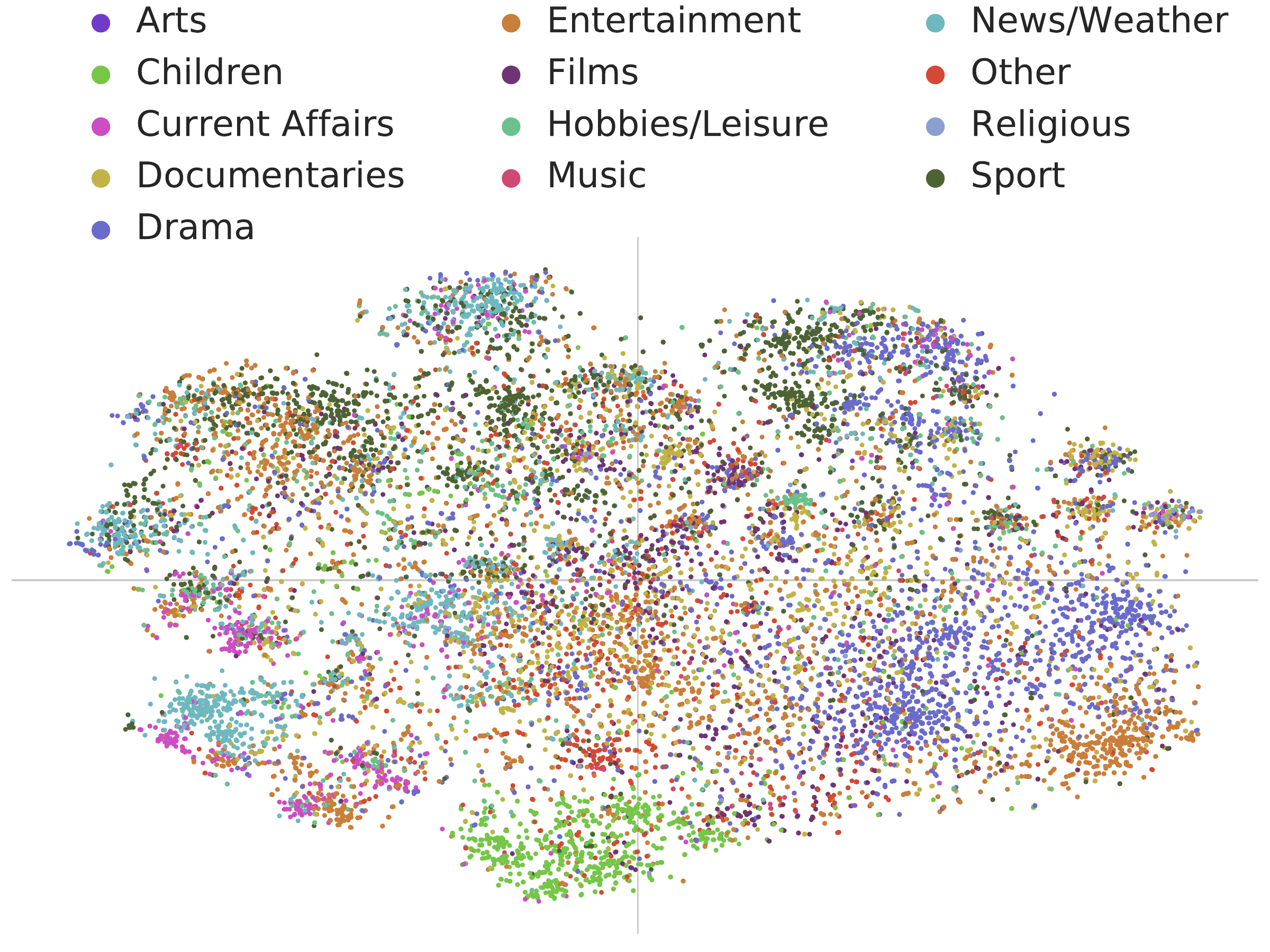}%
    }
    \subfloat[Recommended genre\label{fig:b}]{
        \includegraphics[width=0.33\textwidth]{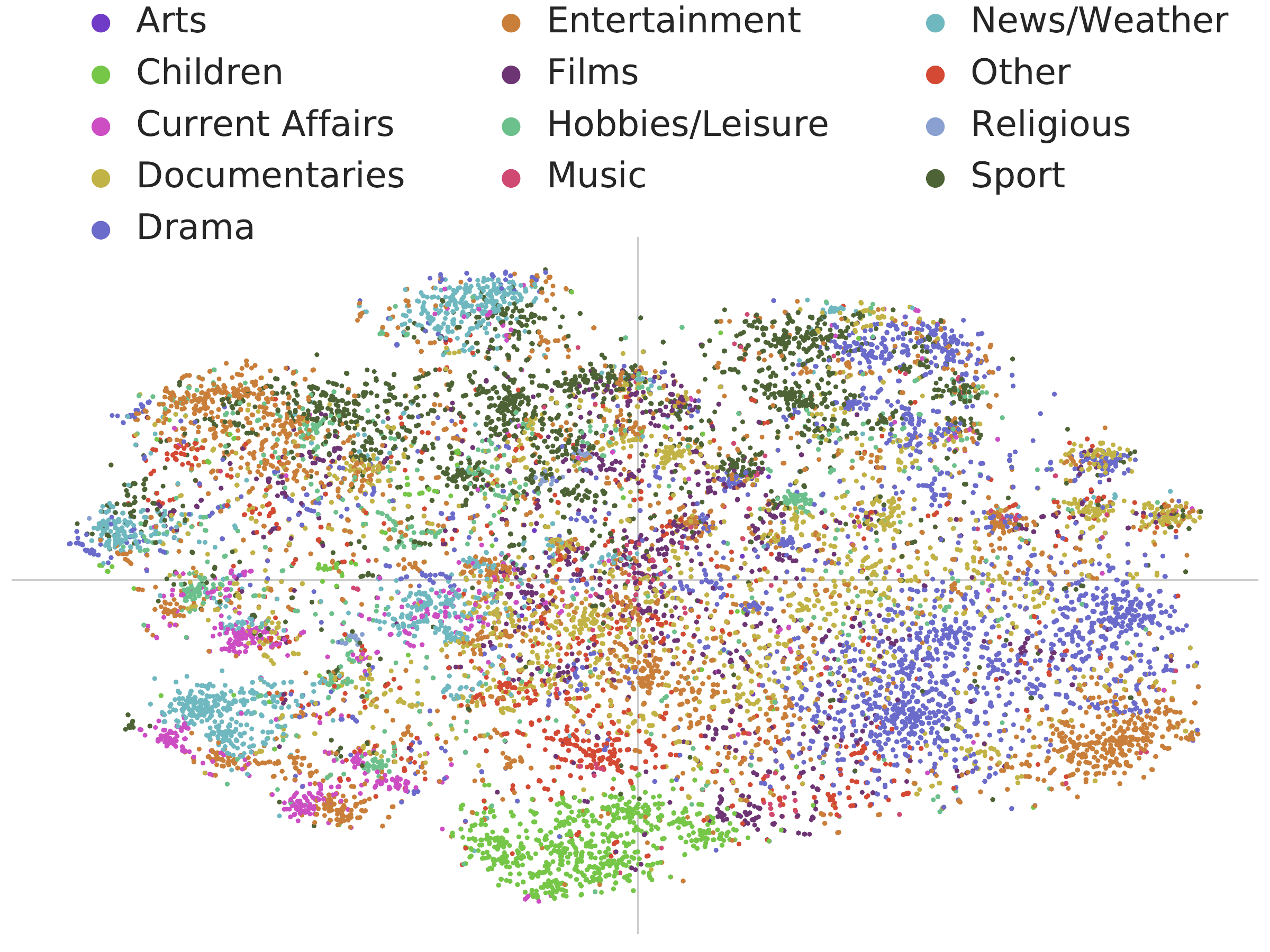}%
    }
    \subfloat[Observed vs. recommended genre\label{fig:c}]{
        \includegraphics[width=0.33\textwidth]{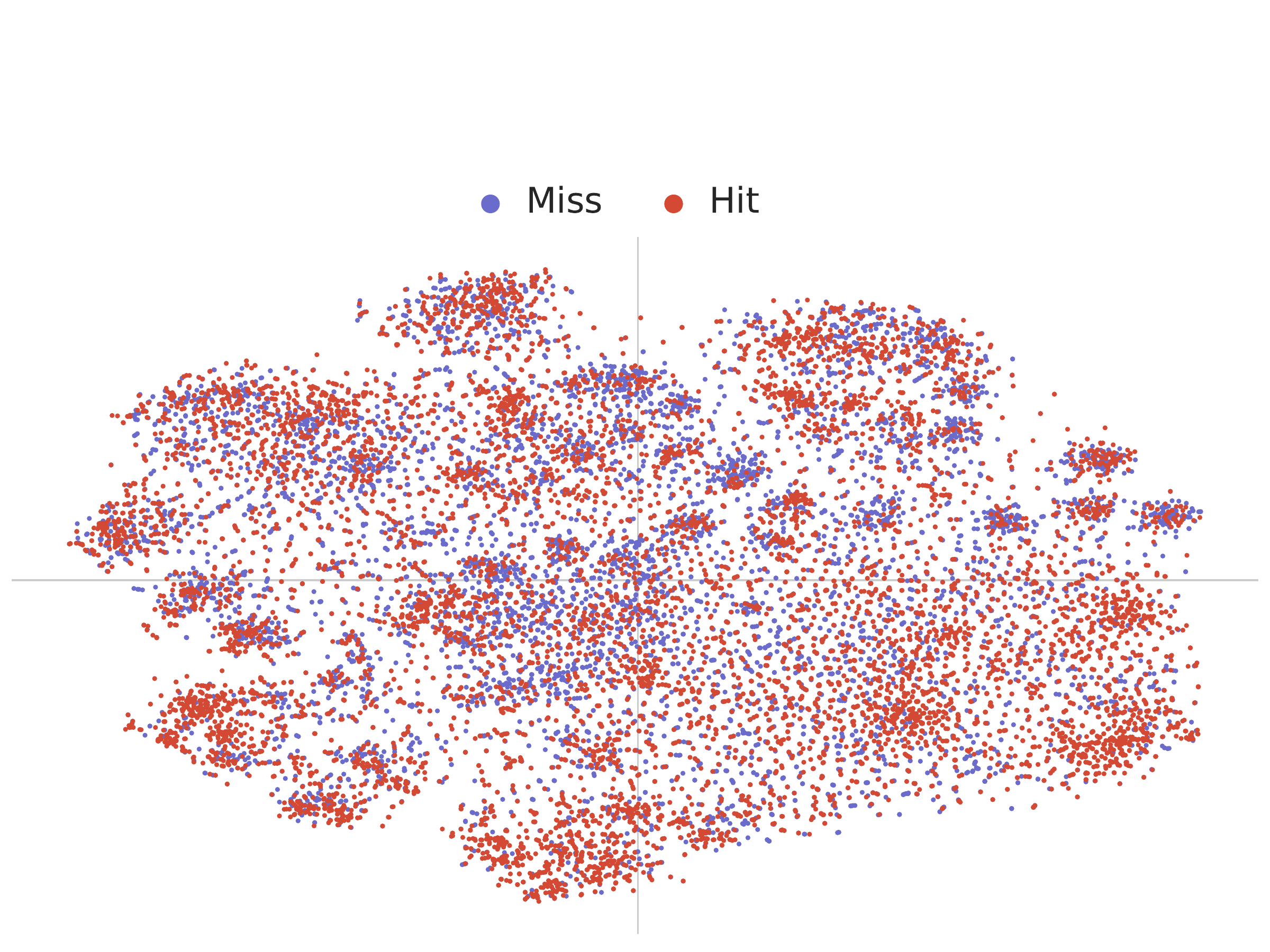}%
    }\\
    \subfloat[Day of week\label{fig:d}]{
        \includegraphics[width=0.33\textwidth]{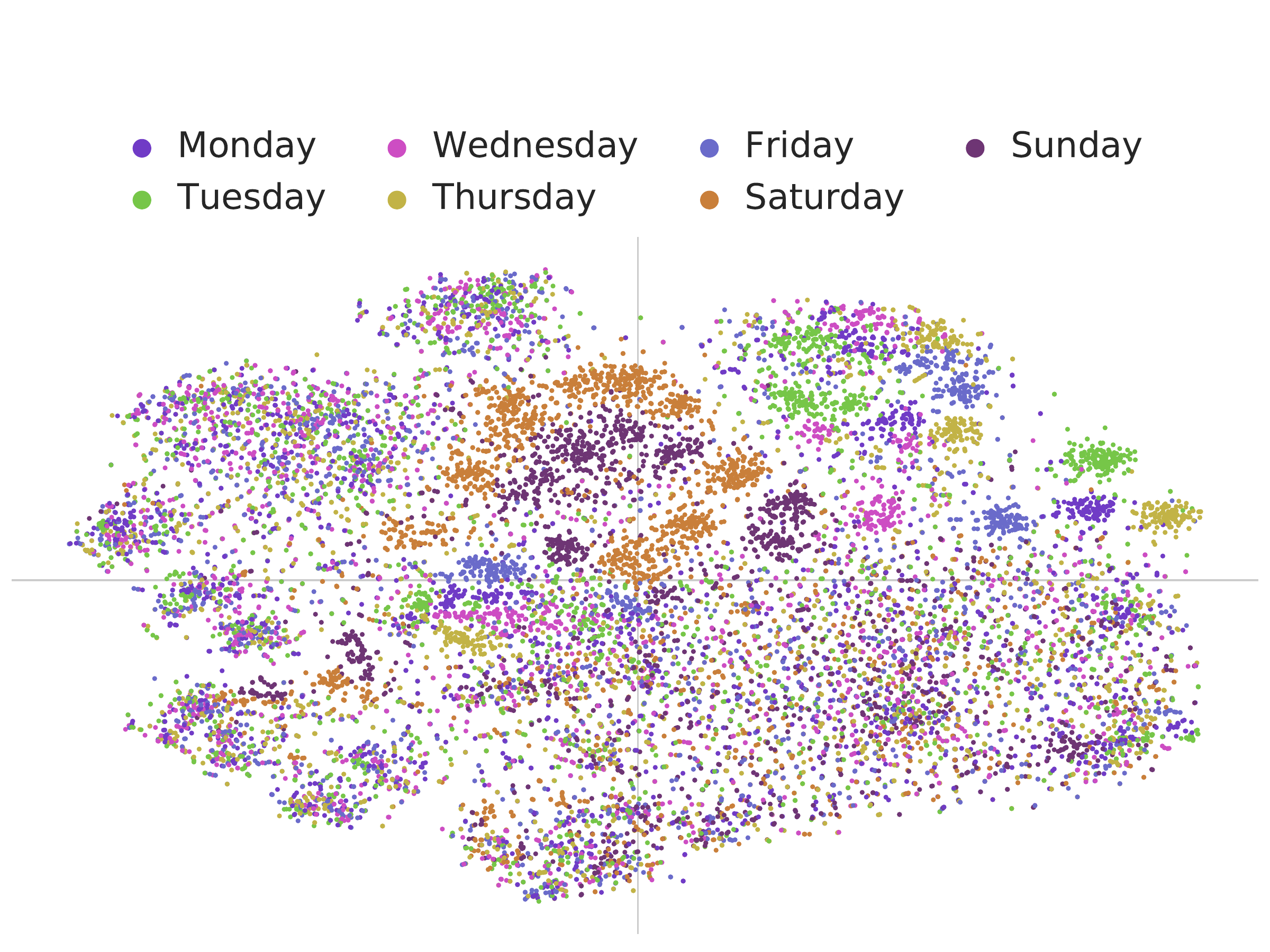}%
    }
    \subfloat[Time of day\label{fig:e}]{
        \includegraphics[width=0.33\textwidth]{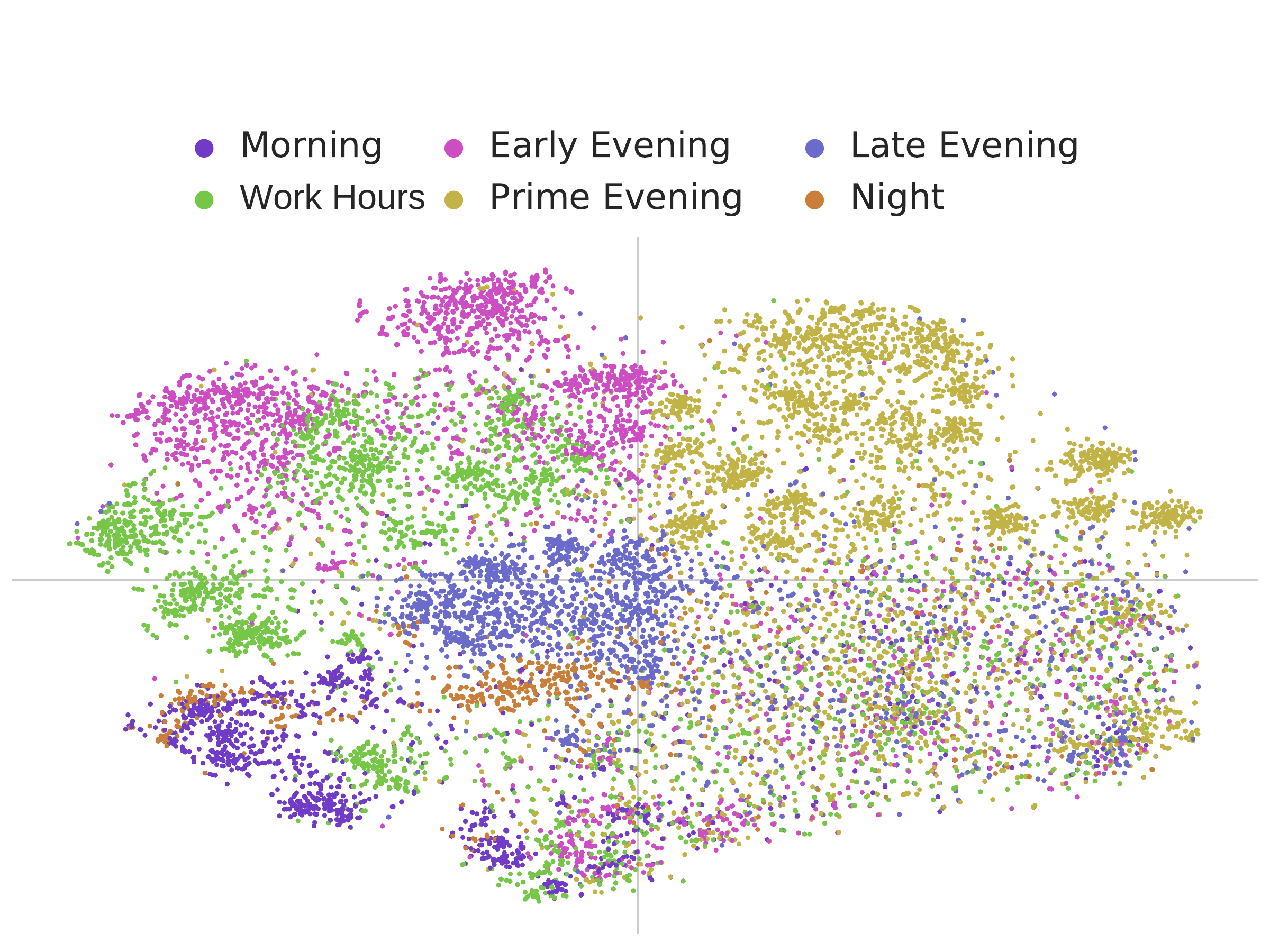}%
    }
    \subfloat[Session activity type\label{fig:f}]{
        \includegraphics[width=0.33\textwidth]{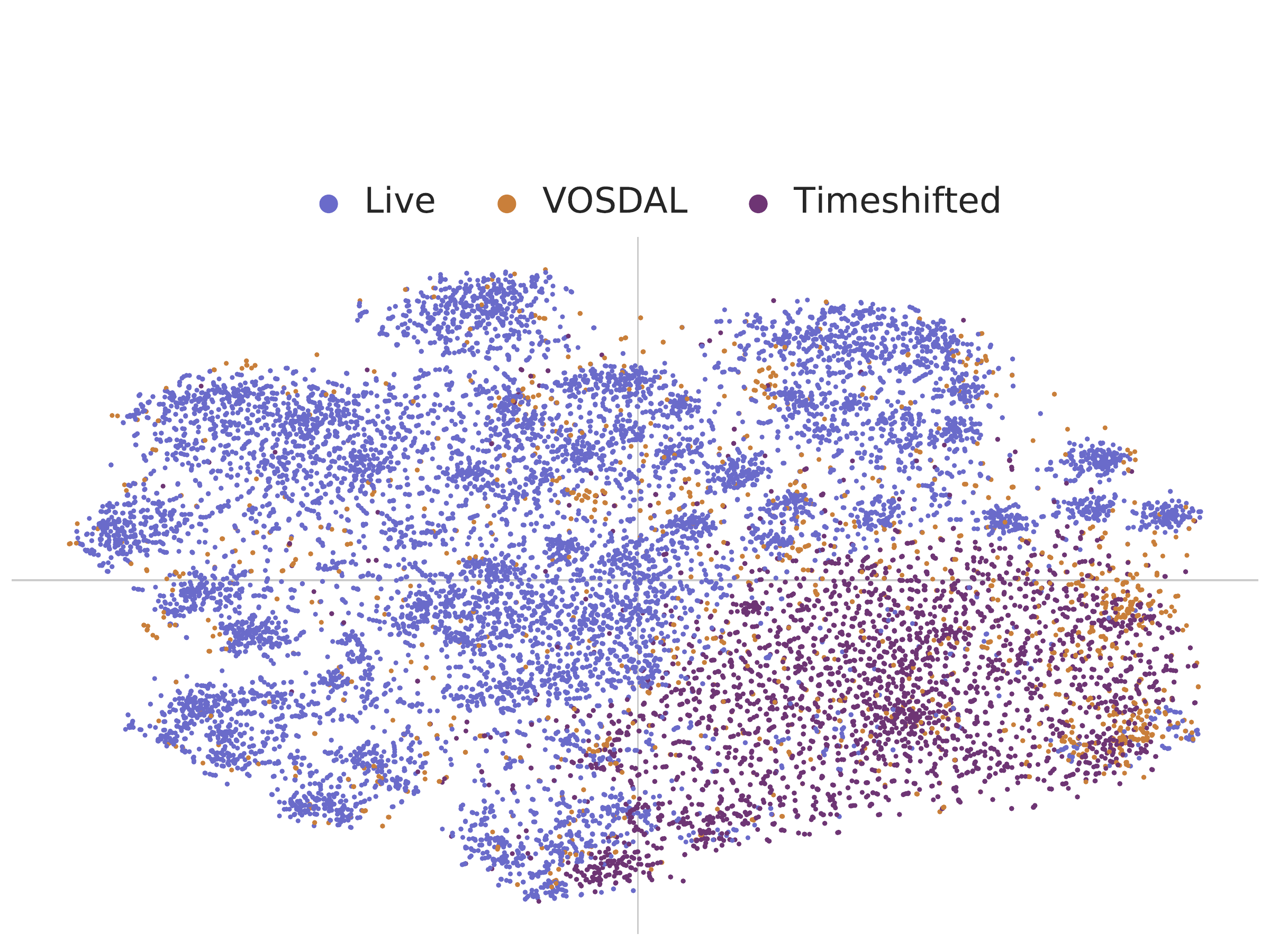}%
    }\\
    \subfloat[Gender distribution\label{fig:g}]{
        \includegraphics[width=0.33\textwidth]{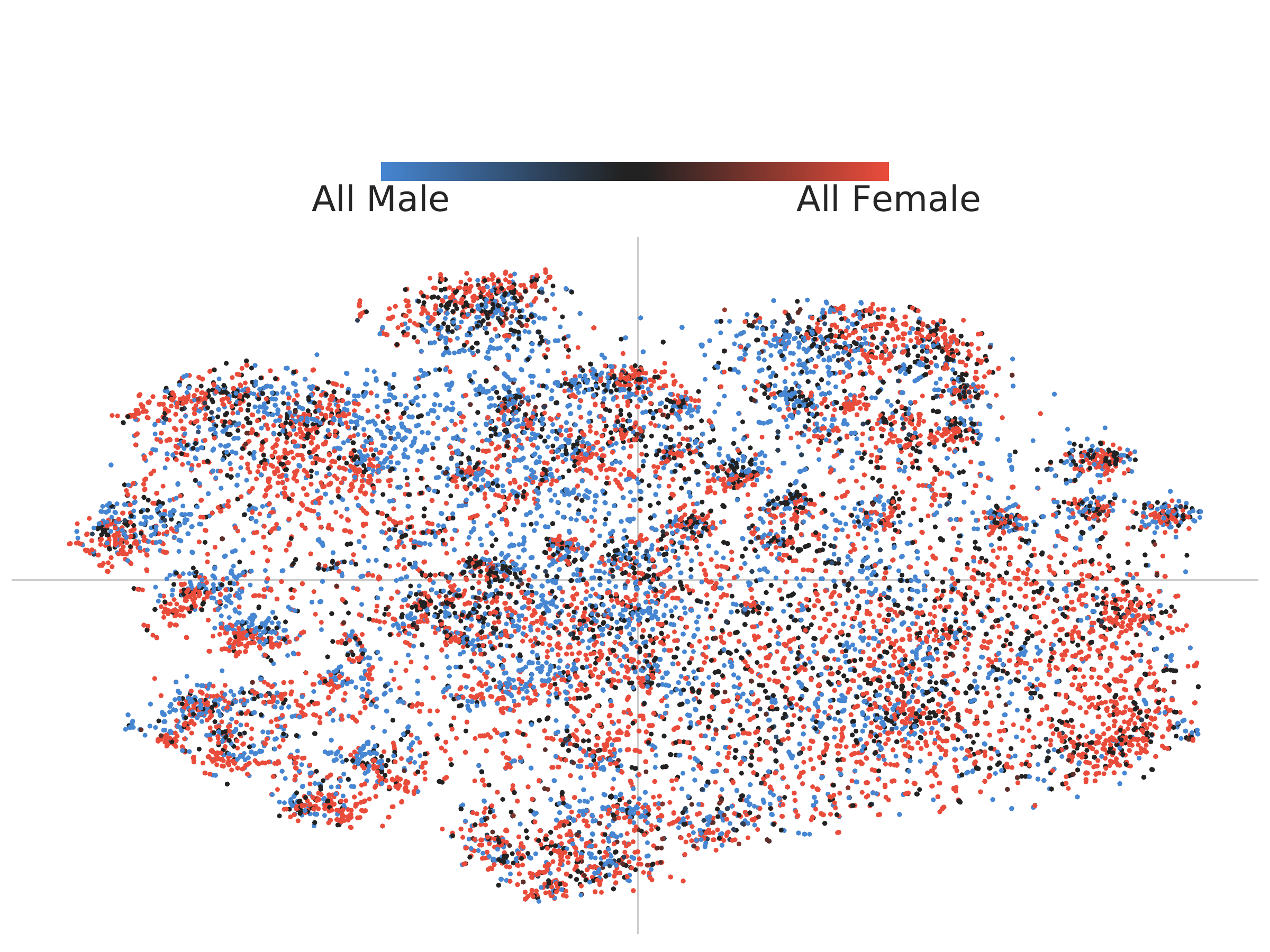}%
    }
    \subfloat[Average age\label{fig:h}]{
        \includegraphics[width=0.33\textwidth]{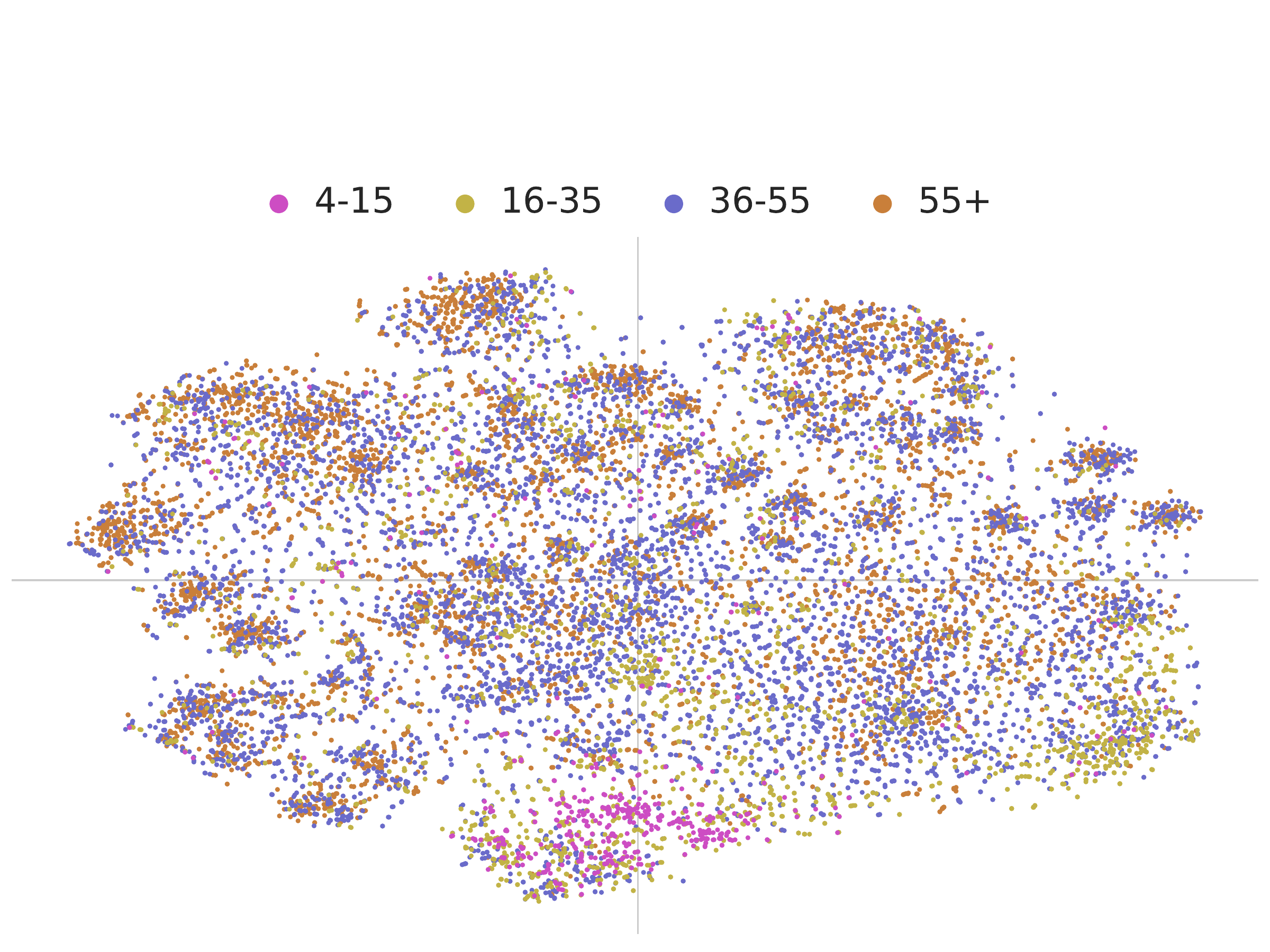}%
    }
    \subfloat[Social\label{fig:i}]{
        \includegraphics[width=0.33\textwidth]{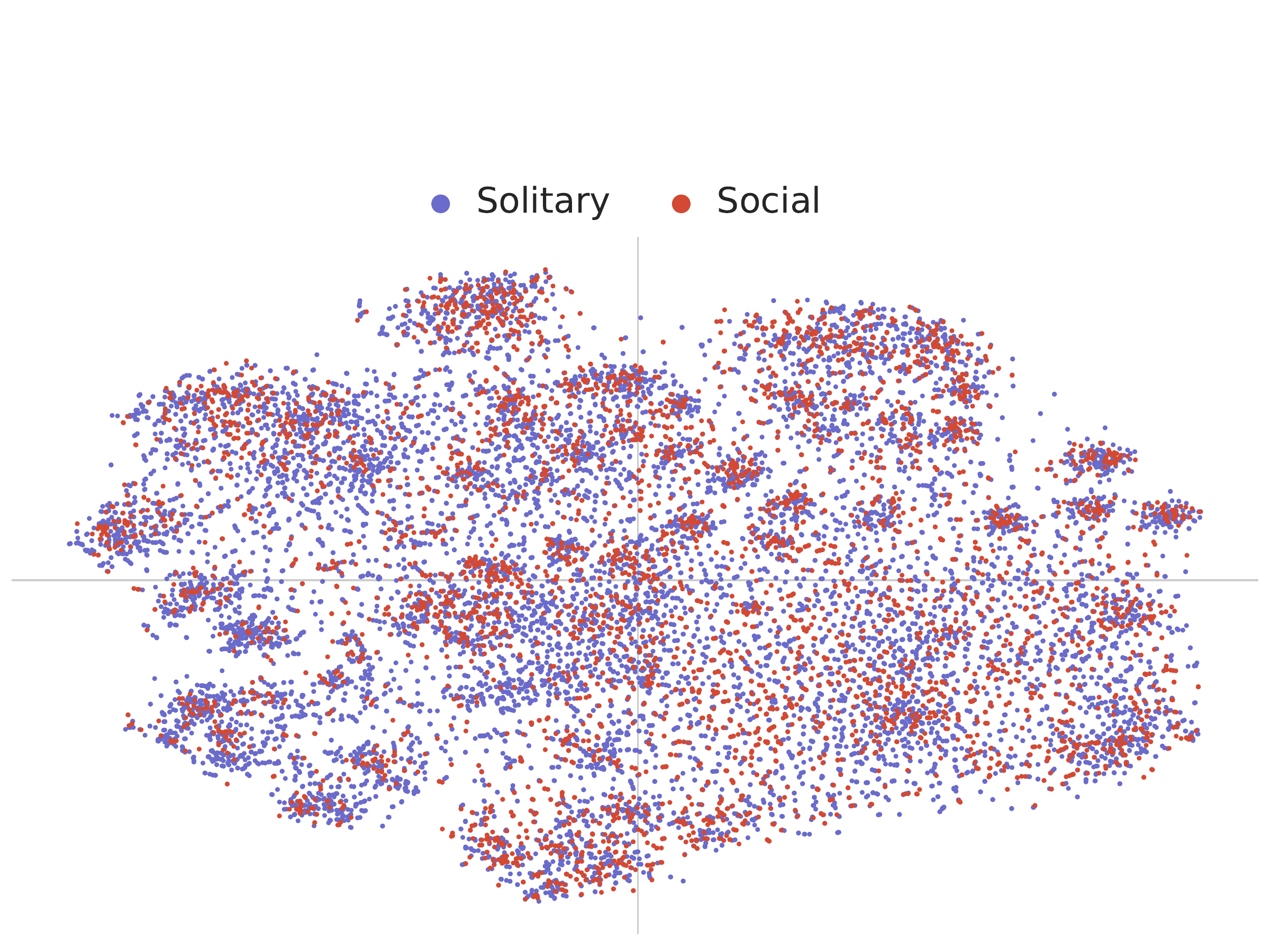}%
    }
    \caption{R-JCCE context embeddings of 20,000 random samples from the BARB test set reduced to two dimensions using t-SNE. Each sub plot show the same embeddings, but color them from different aspects of the viewing events. The colors of (a)-(c) are based on the associated content of each viewing situation, whereas (d)-(i) are colored from viewing context. The figure is zoomable.}
    \label{fig:context_embs}
\end{figure*}
Patterns in viewing context and content are complex and not easily unraveled.
This experiment seeks to uncover some of the underlying structures of television consumption.
To this end, we randomly sample 20,000 viewing events from the test set, such that they are not seen during training.
We then compute their 50-dimensional context embeddings using the context encoder, $\phi_C$, and reduce to two dimensions with t-SNE~\cite{Maaten2008} using cosine similarity.
The embeddings are shown in Fig.~\ref{fig:context_embs}, where each sub plot presents the 20,000 embeddings colored from different variables of the viewing events.\footnote{Note that the figure can not be directly compared to Fig.~\ref{fig:content_emb}, since a new t-SNE mapping is computed and also t-SNE is non-deterministic.}

In Fig.~\ref{fig:a} the context embeddings are colored from the top level genre that is consumed in each viewing event, \ref{fig:b} instead show the recommended top level genre by R-JCCE, and \ref{fig:c} visualizes for each viewing event whether the recommended top level genre matches the observed top level genre.
The figure shows that context embeddings group based on consumed content, in line with Fig.~\ref{fig:content_emb}.
As an example, the \textit{Children} genre clearly stands out at the bottom of the plot.
It is however also evident that some viewing situations are difficult to group in two dimensions in terms of consumed content, since neighboring events consist of a diverse set of genres.
Note that R-JCCE is able to model some of these complex scenarios that are not easily visualized in two dimensions.
Judged from the hits and misses in \ref{fig:c}, the areas with a high degree of overlap tend to be more prone to mistakes.

Figs.~\ref{fig:d} and \ref{fig:e} show the temporal settings of each viewing event.
The time of day is colored according to the following hour scheme. 
0-6: Night, 6-9: Morning, 9-15: Work Hours, 15-18: Early evening, 18-21: Prime evening, 21-24: Late evening. 
The plots confirm that the model uses temporal context as an important feature.
Time of day seems to be the strongest indicator, while the day of the week mainly serves to highlight weekends and distinguish days of prime time viewing.
Note, however, that the fourth quadrant of the visualization is not structured from temporal settings.
The reason can be seen in Fig.~\ref{fig:f} that shows the activity type.
That is, the viewing events in the fourth quadrant mainly consist of timeshifted viewing on same day as live (VOSDAL) as well as timeshifted viewing up to 28 days after the live broadcast (Timeshifted).
For this type of content, the viewers themselves decide when to watch, which has a large impact on the underlying temporal patterns.

Fig.~\ref{fig:g} is colored on a continuous scale from the gender distribution of the viewer(s) in each viewing event.
As an example, a group with two female viewers is all female, a group with one female and one male is half.
\textit{Solitary viewing} maps to either all male or all female.
Fig.~\ref{fig:h} shows the average age of the viewer(s).\footnote{It would be better if it was possible to indicate individual viewers' age.
As an example, a child watching together with a grandparent will have an average age indistinguishable from a solitary middle aged viewer.
This disadvantage only applies to this specific visualization, since the model receives the age of each viewer as input.}
As expected, the \textit{Children} genre cluster exhibits a low age compared to the rest.
Also, Fig.~\ref{fig:i} shows that a large part of the children are co-viewing.

Combining the sub plots, it is now possible to see tendencies such as young females viewing entertainment timeshifted (in the bottom right of the plot), or males viewing prime time Tuesday sport.
Several such patterns can be found, e.g. using clustering, and used to recommend content or understand some of the complex structures of television consumption.

\subsubsection{Entanglement of Embeddings}\label{sec:exp_rep3}
So far the evaluation of the embeddings have been based on qualitative insights.
In this experiment we will present a quantitative measure of entanglement among the embedded viewing events.
The intuition is that a perfect model is able to isolate context embeddings into separate clusters for each content.
We have already seen that in two dimensions R-JCCE has overlap between the genres, cf. Fig.~\ref{fig:a}.

We use soft nearest neighbor measure (SNNM) as the metric, which has previously been used in \cite{Frosst2019a} as a loss function.
Specifically, we define SNNM as:
\begin{align}
    &\text{SNNM} = \nonumber\\
    &-\frac{1}{n}\sum_{i=1}^n \log \frac{\sum_{j\neq i}^n \mathbbm{1}(I_i=I_j)e^{-\theta(\phi_C(C_i),\phi_C(C_j))/T}}{\sum_{k\neq i}^n e^{-\theta(\phi_C(C_i),\phi_C(C_k))/T}},
\end{align}
where $n$ is the number of embeddings, $T$ is the temperature, $\mathbbm 1$ is an indicator function that has a value of one if $I_i$ and $I_j$ match and zero otherwise, and $\theta(x,y)$ is the angular distance between two embeddings:
\begin{equation}
    \theta(x,y) = \frac{1}{\pi} \arccos \frac{x^\mathsf{T}y}{||x||\,||y||}.
\end{equation}
Here we use angular distance instead of cosine similarity, since angular distance is a proper distance metric.

\begin{figure}[tb]
    \centering
    \includegraphics[width=\columnwidth]{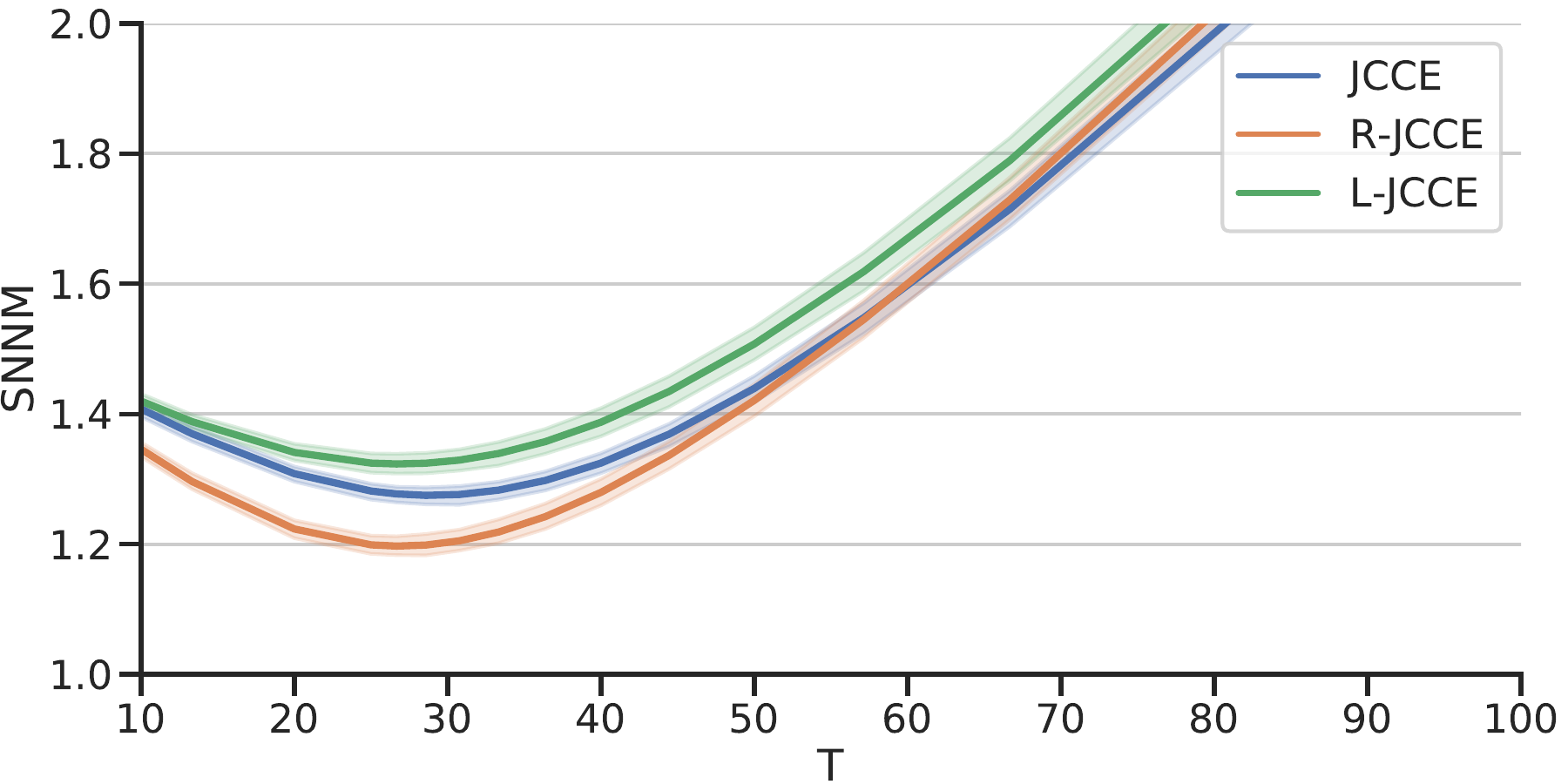}
\caption{SNNM means and 95\% confidence intervals as a function of temperature for the BARB test set. A lower score is better.}
\label{fig:SNNM}
\end{figure}

Fig.~\ref{fig:SNNM} shows the result of running 20 repetitions with random sampling of $n=512$ embeddings from the BARB test set for each of the three JCCE variants.
At the optimal temperature, R-JCCE suffers the least from entangled context embeddings and achieves an average SNNM score of approximately 1.2.

\subsubsection{Context and Content Embedding Similarity}\label{sec:exp_rep4}
\begin{figure*}
    \centering
    \includegraphics[width=1.0\textwidth]{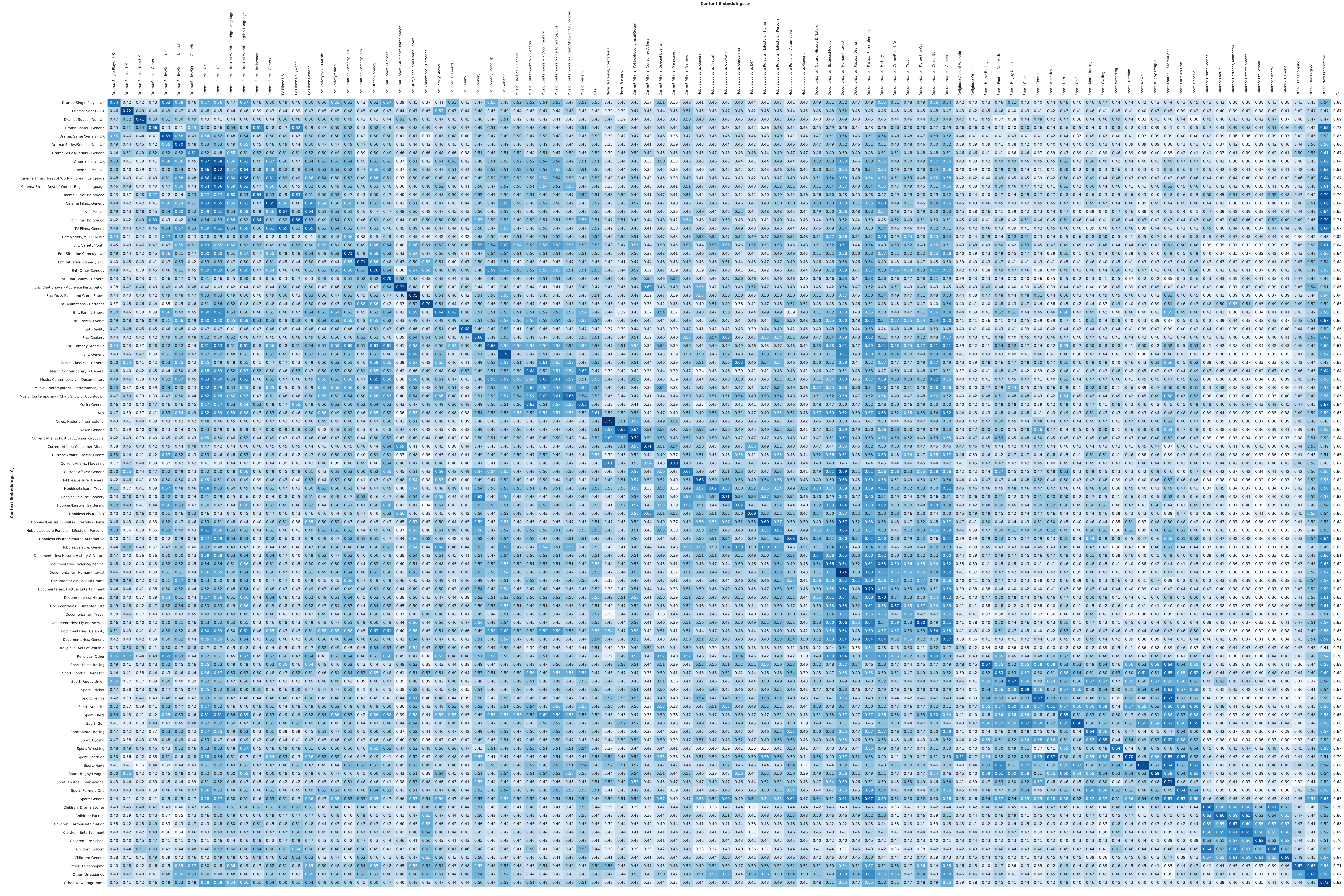}
\caption{Angular similarity between average R-JCCE context embedding associated with each genre in the BARB dataset and the genre content embedding. The figure is zoomable.}
\label{fig:dist_mat}
\end{figure*}
In the last experiment we study in detail the similarity between context and content embeddings for the BARB dataset.
To this end, we divide the BARB test set into subsets, one for each genre, and compute all context embeddings associated with each genre.
For each genre, we then calculate the average context embedding, $\overline \phi_C$:
\begin{equation}
    \overline \phi_{C,i} = \frac{1}{|Y_i|}\sum_{y\in Y_i}\phi_C(C_y),\ i=\{1,\ldots,M\},
\end{equation}
where $Y_i$ is the set of indices of viewing events with content $I_i$.
We then measure the angular similarity between the average context embedding and all $M=94$ genre embeddings: 
\begin{equation}
    1-\theta(\overline\phi_{C,i},\phi_I(I_j)),\ i,j=\{1,\ldots,M\}.
\end{equation}
The resulting similarities are shown in Fig.~\ref{fig:dist_mat} with average context embeddings along the rows and content embeddings along the columns.
The right most column lists the variance of context embeddings compared to their average context embedding:
\begin{equation}
    \sigma_{C,i} = \frac{1}{|Y_i|}\sum_{y\in Y_i}1-\theta(\overline\phi_{C,i},\phi_C(C_y)),\ i=\{1,\ldots,M\}.
\end{equation}

The results in Fig.~\ref{fig:dist_mat} provide a detailed look into some of the aspects touched upon in Section~\ref{sec:exp_rep1}.
The \textit{Sport} and \textit{Children} top level genres stand out as the most separable clusters.
\textit{Documentaries} have a vertical pattern, which indicates that the genre generally achieves high similarity scores and could be prone to cause false recommendations.
Note that for most genres the diagonal of the similarity matrix is the preferred genre. 
That is, the context embeddings tend to fall close to the associated content embedding, proving the value of using R-JCCE to learn the shared latent space.

\section{Concluding Remarks}\label{sec:con}
In this work, we explored deep embeddings learned jointly for context and content in the television domain.
We introduced the unified framework of JCCE that delivers context-aware recommendations, while also supplying tools for exploring patterns of context-content, context-context, and content-content relationships.
The $N$-pairs learning objective of the original JCCE model was relaxed to improve performance further in terms of recommendations with very few suggestions.
We demonstrated the capability of JCCE by achieving state-of-the-art performance for recommendations in the television domain on both a proprietary and a publicly available dataset.
We furthermore presented qualitative and quantitative experiments evaluating the representation capabilities of JCCE, e.g. by visualizing learned structures in the shared latent space.

Since there is a tendency in multimedia consumption to move towards on-demand based viewing, and our results show that temporal patterns are weakened for timeshifted viewing, it will be of interest to study this issue further in the future.
It may also be of interest to explore the relation between genre and item based recommendations, and how genres can potentially be used to reduce the search space.

\ifCLASSOPTIONcompsoc
  \section*{Acknowledgments}
\else
  \section*{Acknowledgment}
\fi

The authors would like to thank Vinoba Vinayagamoorthy, Jacob L. Wieland, and the BBC for kindly hosting Miklas S. Kristoffersen during parts of the work, and for providing guidance and access to data.



\ifCLASSOPTIONcaptionsoff
  \newpage
\fi

\balance



\bibliographystyle{IEEEtran}
%
%
%

\begin{IEEEbiography}[{\includegraphics[width=1in,height=1.25in,clip,keepaspectratio]{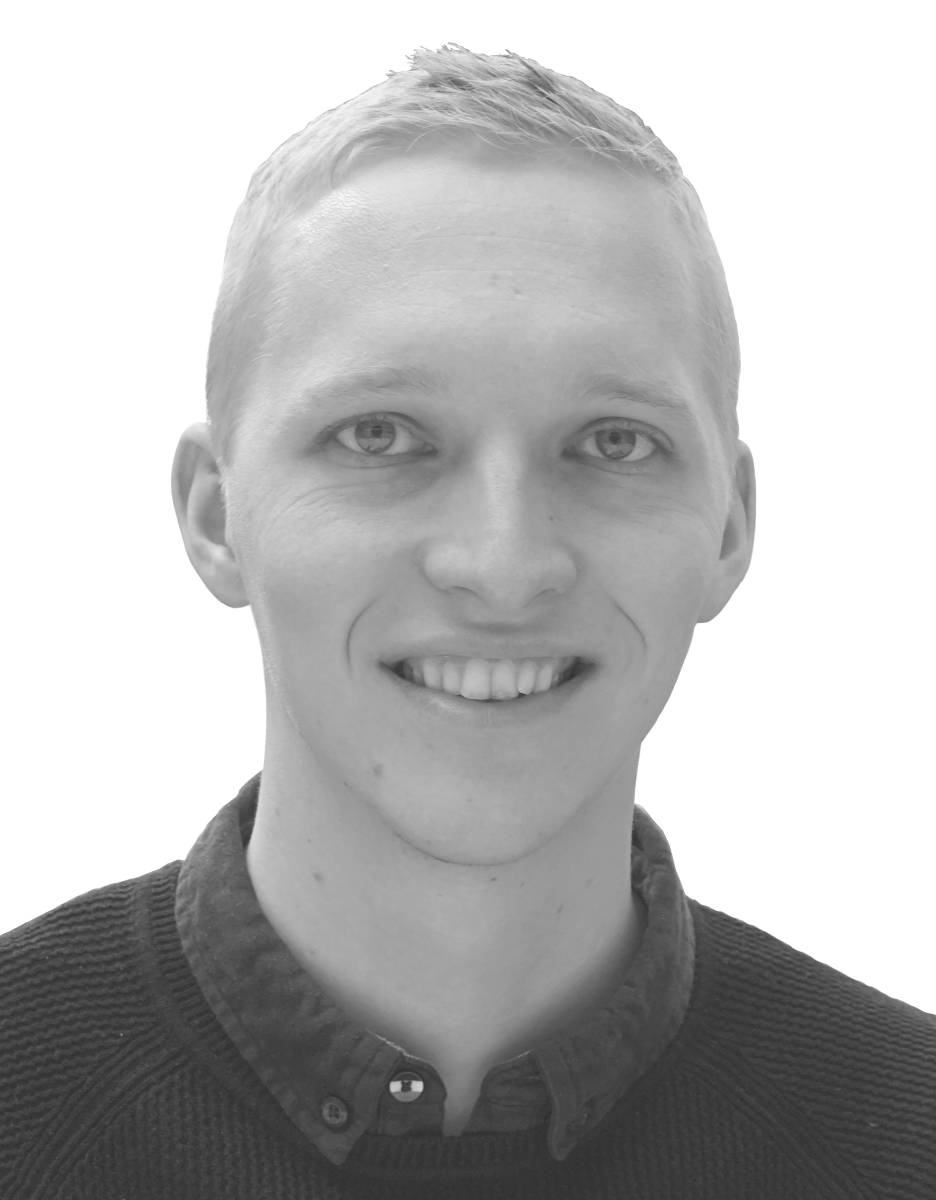}}]{Miklas Str{\o}m Kristoffersen}
received the B.Sc. and M.Sc. degrees in electrical engineering with specialization in vision, graphics, and interactive systems from Aalborg University, Aalborg, Denmark, in 2014 and 2016, respectively. In 2020, he received the industrial Ph.D. degree from Aalborg University and Bang \& Olufsen A/S. He is currently employed at Salling Group A/S in Denmark. His research interests include machine learning, context-aware recommender systems, and self-supervised learning.
\end{IEEEbiography}

\begin{IEEEbiography}[{\includegraphics[width=1in,height=1.25in,clip,keepaspectratio]{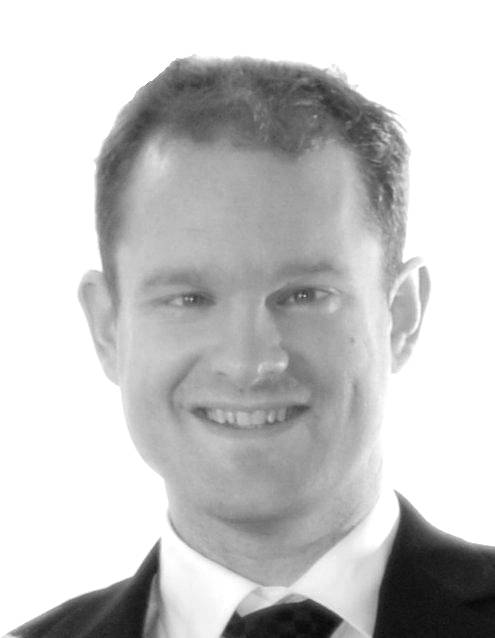}}]{Sven Ewan Shepstone}
(M'11)
received the B.S. and M.S. degrees in Electrical Engineering from the University of Cape Town in 1999 and 2002 respectively. From 2005-2010 he worked as a Systems Engineer in the field of Ethernet and broadband communications for Ericsson A/S in Denmark, and has been employed at Bang \& Olufsen A/S in Denmark since 2010. In 2015 he received the Ph.D. degree from Aalborg University. His research interests include frameworks for digital TV and the application of  speech technologies to recommender systems. He was the recipient of the IEEE Ganesh N. Ramaswamy Memorial Student Grant at ICASSP 2015.

%

\end{IEEEbiography}

\begin{IEEEbiography}[{\includegraphics[width=1in,height=1.25in,clip,keepaspectratio]{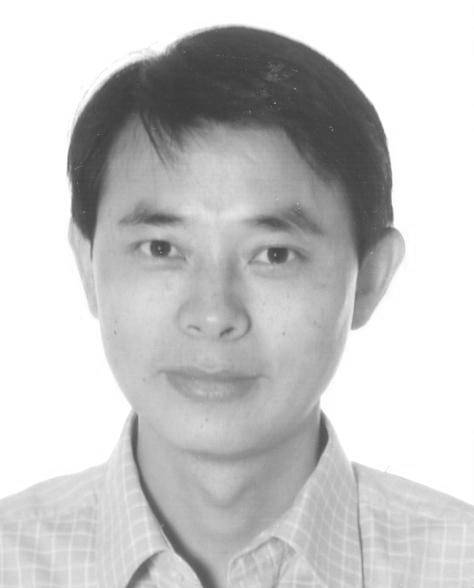}}]{Zheng-Hua Tan}

(M'00--SM'06)
received the B.Sc. and M.Sc. degrees in electrical engineering from Hunan University, Changsha, China, in 1990 and 1996, respectively, and the Ph.D. degree in electronic engineering from Shanghai Jiao Tong University (SJTU), Shanghai, China, in 1999.

He is a Professor in the Department of Electronic Systems and a Co-Head of the Centre for Acoustic Signal Processing Research at Aalborg University, Aalborg, Denmark. He is also a Co-Lead of the Pioneer Centre for AI, Denmark. He was a Visiting Scientist at the Computer Science and Artificial Intelligence Laboratory, MIT, Cambridge, USA, an Associate Professor at SJTU, Shanghai, China, and a postdoctoral fellow at KAIST, Daejeon, Korea. His research interests include machine learning, deep learning, pattern recognition, speech and speaker recognition, noise-robust speech processing, and multimodal signal processing. He has authored/coauthored over 200 refereed publications. He was the Chair of the IEEE Signal Processing Society Machine Learning for Signal Processing Technical Committee (MLSP TC) and an Associate Editor for the IEEE/ACM TRANSACTIONS ON AUDIO, SPEECH AND LANGUAGE PROCESSING. He has served as an Editorial Board Member for Computer Speech and Language and was a Guest Editor for the IEEE JOURNAL OF SELECTED TOPICS IN SIGNAL PROCESSING and Neurocomputing. He was the General Chair for IEEE MLSP 2018 and a TPC Co-Chair for IEEE SLT 2016.

\end{IEEEbiography}

\end{document}